\documentclass[sigconf]{acmart}
\AtBeginDocument{%
  }

\author{Kyzyl Monteiro}
\orcid{0000-0002-2723-9500}
\affiliation{%
  \institution{Carnegie Mellon University}
  \city{Pittsburgh}
  \state{PA}
  \country{USA}
}
\email{kyzyl@cmu.edu}

\author{Minjung Park}
\affiliation{%
  \institution{Carnegie Mellon University}
  \city{Pittsburgh}
  \state{PA}
  \country{USA}
}
\email{mpark2@andrew.cmu.edu}

\author{Alexander Ioffrida}
\affiliation{%
  \institution{New York University}
  \city{New York}
  \state{NY}
  \country{USA}
}
\email{adf7388@nyu.edu}

\author{Angelina Sanna}
\affiliation{%
  \institution{Carnegie Mellon University}
  \city{Pittsburgh}
  \state{PA}
  \country{USA}
}
\email{asanna@andrew.cmu.edu}

\author{Hao-Ping (Hank) Lee}
\affiliation{%
  \institution{Carnegie Mellon University}
  \city{Pittsburgh}
  \state{PA}
  \country{USA}
}
\email{haopingl@andrew.cmu.edu}

\author{Niloofar Mireshghallah}
\affiliation{%
  \institution{Carnegie Mellon University}
  \city{Pittsburgh}
  \state{PA}
  \country{USA}
}
\email{niloofar@cmu.edu}

\author{Yang Wang}
\affiliation{%
  \institution{University of Illinois at Urbana-Champaign}
  \city{Champaign}
  \state{IL}
  \country{USA}
}
\email{yvw@illinois.edu}

\author{Sauvik Das}
\orcid{0000-0002-9073-8054}
\affiliation{%
  \institution{Carnegie Mellon University}
  \city{Pittsburgh}
  \state{PA}
  \country{USA}
}
\email{sauvik@cmu.edu}

\usepackage{tabularx}
\usepackage{graphicx}
\graphicspath{{figures/}}

\acmConference[Preprint]{Preprint}{}{}
\renewcommand{\footnotetextcopyrightpermission}[1]{%
  \footnotetext{
    Preprint.
  }
}

\begin{document}

\title{When Are LLM Inferences Acceptable? User Reactions and Control Preferences for Inferred Personal Information}



\renewcommand{\shortauthors}{}

\begin{abstract}

Ask ChatGPT about vacation planning, and it may infer your income. Ask it about medication, and it may infer your medical history. Because such inferences can expose more information than users intend to reveal, prior work argues that they are a defining privacy risk of LLM-based systems. Yet prior work has mostly shown that LLMs \emph{can} make potentially violating inferences, not how users experience those inferences nor what controls users may want governing their use. We built the Reflective Layer, a visualization tool that surfaces example unstated inferences from users’ own ChatGPT histories, and used it in a mixed-methods study with 18 regular ChatGPT users evaluating 215 surfaced inferences from their own conversations. Counterintuitively, participants reacted more strongly with curiosity and interest rather than distress and concern. Discomfort arose mainly when inferences felt misrepresentative of the user or misaligned with expected use. Participants were also markedly less comfortable with advertisers and third-party applications using those inferences than with platform providers. These findings suggest that the acceptability of LLM inferences is governed not only by its content, but by context-sensitive norms around how they are generated, retained within the platform, and transmitted beyond it.
\end{abstract}



\maketitle

\section{Introduction}


People share deeply personal information with large language model (LLM)-based systems~\cite{mireshghallah2024trust, zhang2024s}.
Moreover, from these disclosures, LLMs can infer much more about users than what they explicitly state~\cite{staab2023beyond,shaikh2025creating,tomekcce2024private}.
For example, if Alice asks an LLM to help draft an email to her manager requesting sick leave, that same LLM, appropriately prompted, can infer her workplace role, relationship with her manager, and aspects of her health.
This ability to make personal inferences---about users' intents, preferences, circumstances, and needs beyond what they have shared---is a defining privacy risk of LLM-based systems that goes beyond verbatim memorization of conversational data~\cite{staab2023beyond, mireshghallah2025position,lee2024deepfakes}.
The dangers of these inference-based privacy risks are even more salient as prominent model providers, like OpenAI, have introduced personalized advertisements based on conversation context into these conversational assistants~\cite{openai_chatgpt_ads}, are exposing tool actions by third-party vendors through the Model Context Protocol~\cite{anthropic_mcp, openai_mcp}, and as users delegate tasks to LLM-powered agentic harnesses that can autonomously act on, use, or share user information~\cite{openclaw}.

Simultaneously, this same inference capability is lauded and being further developed for its utility. The AI and HCI communities are moving towards an aspirational vision of ``proactive'' agents~\cite{maes1993learning, lieberman1995letizia, shneiderman1997direct, lam2026just, shaikh2026learning} that predict the user's next action based on what can be inferred about the user's context and goals from their conversations and activities. One instantiation of this vision is the \emph{general user model} (GUM)~\cite{shaikh2025creating}---a computational model built from conversational and activity data that makes deep inferences about user intent, context, and need in order to personalize AI behavior. These proactive agents promise to reduce user prompting burden through inferred grounding: they use inferences to ``understand'' a user's context without the user needing to explicitly share it.

These complementary perspectives---on the utility versus intrusiveness of LLM inferences---are now coming to a head.
As agentic AI (e.g., Claude Desktop, OpenClaw) rises in popularity, the risks and benefits of personal inferences will increasingly shape users' day-to-day experience with LLM-powered systems.

But when do users find these inferences acceptable, and when do they find them intrusive?
To date, work on revealing and managing LLM privacy risks for users has focused on \emph{memory}---what conversational LLMs save and retrieve from prior conversations~\cite{chen2026relational, zhang2025understanding, mekioussa2026towards, zhang2024ghost}---rather than on \emph{inference}: i.e., what LLMs can learn about users that they did not explicitly state or share.
Without the user perspective on inferences, it is difficult to assess what controls are necessary, or how to design for inferences that benefit users in ways they find acceptable, not intrusive.

To bridge this gap, we pose two research questions. First, \textit{how do users feel about and evaluate the personal inferences that an LLM can make about them from their conversation history} (RQ1)? Second, \textit{what access-control preferences do users hold for such inferences, and what reasoning underlies those preferences} (RQ2)?

A critical challenge with answering these questions is that LLM inferences remain opaque to users. Because users rarely see what \emph{is being or could be} inferred about them, it is difficult for them to assess or articulate their reactions.
We built the \textit{Reflective Layer}: a research probe that surfaces inferences derived from users' own chat histories. The Reflective Layer adapts the GUM pipeline introduced in prior work~\cite{shaikh2025creating} to conversational data exported from ChatGPT. GUM already distinguishes \emph{observations} (factual records) from \emph{propositions} (confidence-weighted inferences about users), and surfaces the latter for downstream AI assistants to consume silently. We adapt this architecture to build a user-facing visualization tool: adding an explicit observation-extraction step so that each inference traces back through its supporting observations to the specific chat messages that produced it, organizing inferences into categories drawn from user modeling theory~\cite{brusilovsky2007user, kobsa2007generic}, and scoring sensitivity via continuous Contextual Integrity (CI) judgments~\cite{nissenbaum2004privacy, lan2025contextual}.

With the Reflective Layer as our research instrument, we conducted an eighteen-participant, mixed-methods study in which participants encountered inferences drawn from their own ChatGPT histories. Participants then evaluated these inferences on affective response, relevance, usefulness, and intrusiveness (RQ1). To probe access-control preferences (RQ2), participants additionally rated their preferences across three recipient-and-use contexts (conversational improvement, advertising, and third-party action-taking) and, through a sketching activity, externalized the boundaries they would draw around how these inferences should be accessed.

We found that, on average, participants responded to these inferences more out of curiosity and interest rather than distress and concern, but also that most participants identified at least one inference that they found uncomfortable.
This discomfort arose in two ways: \emph{misrepresentation} (e.g., the user felt that the inference was wrong or presented a distorted view of them) and \emph{misalignment with expected use} (e.g., the inference was correct but the user wanted it hidden or deemed it not worth the system knowing).
Participants varied in their responses to these uncomfortable inferences: some users wanted incorrect or misleading inferences corrected or removed, while others were comfortable with an accurate inference existing but wanted to restrict who else could use it. Finally, we found that participants---even those with high self-reported privacy concern---were generally comfortable with model providers making and using personal inferences, but were less comfortable with advertisers and third-party apps accessing those inferences.

In short, the acceptability of LLM inferences is determined not only by their content but also by context-sensitive norms around how they are \emph{generated}, \emph{retained within the platform}, and \emph{transmitted beyond it}. Moreover, users' preferences for governing these inferences vary across these three stages of the inference lifecycle (from denying inferences from being formed, to correcting them, to restricting how they are used). In sum, we contribute:

\begin{itemize}
    \item \textbf{An empirical account} of how users evaluate and affectively respond to surfaced LLM inferences drawn from their own ChatGPT histories.
    \item \textbf{A characterization} of users' access-control preferences over LLM-derived inferences and the context-sensitive reasoning behind them, with implications for governance at generation, retention, and transmission.
    \item \textbf{The Reflective Layer}, a research probe that surfaces LLM personal inferences from users' own ChatGPT histories for direct user inspection, adapting prior user modeling work~\cite{shaikh2025creating} with CI-based sensitivity scoring~\cite{lan2025contextual}.
\end{itemize}

\section{Related Work}

We are informed by two bodies of prior work. The first establishes that LLMs can derive personal attributes from conversational or activity traces, making inference both central to personalization and an emerging privacy risk. The second, from usable privacy and security, explores how to elicit user reactions to and control preferences for opaque algorithmic systems.

\subsection{LLM Inference Capabilities and Privacy Risks}

\textbf{LLM Inference for Proactive Personalization.} Deep personalization of LLM-powered agents is a guiding goal for much of the AI and HCI communities. Recent work organizes personalized agents around profile modeling, memory, planning, and action execution, showing how user-specific signals can condition how agents interpret requests, retrieve context, form plans, and act~\cite{xu2026toward,fisher2026response}. Empirical work on user profiles similarly shows that historical user records improve personalized LLM outputs across tasks and begins to unpack what profile content contributes to those gains~\cite{wu2024understanding}. Recent systems, informed by the user-modeling literature, have further shown that LLMs can convert observations of user activity into inferences that can be used to predict downstream user actions and goals~\cite{shaikh2025creating, lam2026just, shaikh2026learning}. 

\textbf{LLM Inference as a Privacy Risk.} The source material for such personal inferences is abundant in everyday LLM interactions: conversations routinely contain personal and sensitive content that, when combined with other disclosures, may reveal things about users they do not expect

~\cite{mireshghallah2024trust}. The S\&P community, in particular, has been vocal about the risks entailed by the inferential capabilities of LLMs, showing that models can infer sensitive attributes from users' text and image inputs~\cite{staab2023beyond, tomekcce2024private}. Work on free-form chatbot interactions extends this concern to psychometric traits, such as the Big Five personality inventory~\cite{peters2024large}. This personal attribute leakage can further accumulate across repeated interactions~\cite{mireshghallah2025cimemories}. Existing data-centric protections, including memorization controls, PII redaction, and differential privacy, cannot fully address these risks~\cite{mireshghallah2023can, mireshghallah2025position}, which include contextual-integrity failures and aggregation harms that can arise even without verbatim leakage.

Together, these threads of prior work establish inference as technically plausible, useful, \emph{and} privacy-concerning, but they do not show how users evaluate such inferences. When would users find the use of these inferences for personalization acceptable? When would they find it intrusive? Our work bridges this gap by asking how users' concerns unfold and shape acceptability when personal information is inferred from their own conversational histories.

\subsection{User Perceptions of Inference-driven Profiling}

There is a rich tradition of prior work showing users how algorithms might ``see'' them based on their digital footprints. We drew on this literature as a methodological precedent for our work.

\textbf{Profiling and Behavioral Advertising.} How users experience algorithmic inference has been studied for more than a decade, especially in the context of utility-driven tracking on the web~\cite{englehardt2016online, schaub2016watching, leon2013matters}. Prior work has surfaced activity-based profile inferences in web tracking and online behavioral advertising, in Google and Facebook ad-preference dashboards, and in social-media profiling interfaces, finding that users were often surprised by what had been inferred and used profile visibility to reason about or modify profile entries~\cite{weinshel2019oh, barbosa2021design, kacsmar2023comprehension}. These inferred profiles were experienced as useful or beneficial, but also as creepy, privacy-invasive, surprising, or unsettling~\cite{ur2012smart, reitinger2024does, farke2023does, buchi2023making, hautea2020s}. Together, these studies show the importance of surfacing user-profile inferences to understand users' perspectives and the acceptability of such capabilities. They also provide an important methodological precedent: making hidden profiling legible can elicit users' perceptions of and acceptability judgments about the inferences they encounter, informing systems better aligned with user expectations.

\textbf{Inference Norms in Conversational Systems.} As the threat landscape has expanded beyond web tracking, studies of voice assistants and LLM-based chatbots have examined how users evaluate conversational data use.

These works show that users' judgments depend on factors such as sensitivity, consent, anonymization, and perceived control~\cite{gomez2023sensitive, zhang2024s, tran2025understanding}. Recent AI-mediated studies extend this concern to value extraction, personalization attributes, and LLM-generated personas or interest segments~\cite{yun2026ai, asthana2024know, kronhardt2025all}. Concurrent human-centered audit work has also explored surfacing, to users, attributes that off-the-shelf LLMs associate with their names~\cite{staufer2026human}. As conversational LLM agents become pervasive, personalized, and action-capable, these strands of work motivate asking how users evaluate personal inferences in LLM-driven settings.

\textbf{Memory, User Perceptions, and Inference.} A closely related line of work examines the role of memory in conversational LLMs. Recent user-centered studies examine how users perceive LLM memory, including its relational benefits, privacy strains, and mental-model gaps around what conversational agents retain and reuse~\cite{chen2026relational, zhang2025understanding, mekioussa2026towards}. Building on this work, systems research treats memory as a site of privacy leakage, for example by enabling users to detect and resolve sensitive information stored in memory~\cite{zhang2024ghost}. Similar interface affordances appear in modern LLM platforms' memory interfaces, where users can read and delete stored memories about themselves~\cite{chatgpt_memory,claude_memory,gemini_memory}. Memory, however, is different from inference---inference goes beyond what users explicitly state and aims to elicit conclusions about the user's identity, interests, goals, etc. (see Section~\ref{sec:memory_and_inference}). Wang et al.~\cite{wang2025beyond} show that users struggle to anticipate what attributes an LLM can infer from text and recommend rewriting text to block some inferences, motivating inference-aware data minimization and rewriting support.

We extend this lineage of work by using a similar method---visualizing otherwise opaque algorithmic inferences to users---to model how users perceive the acceptability and intrusiveness of LLM-enabled inferences.

\begin{figure*}[t]
  \centering
  \includegraphics[width=\linewidth]{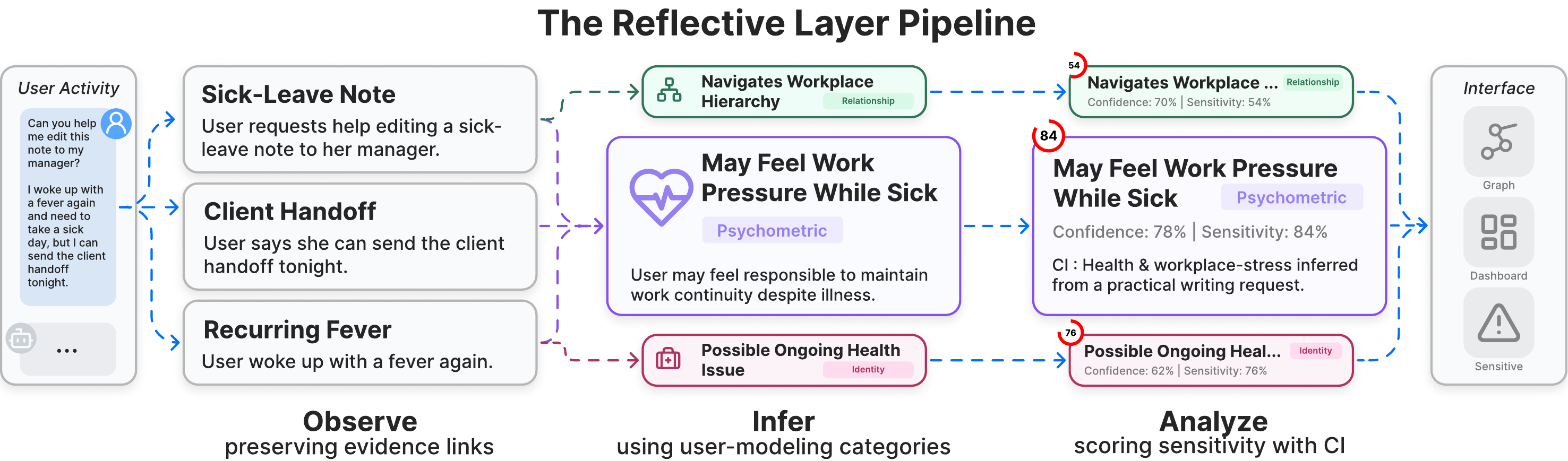}
  \caption{The Reflective Layer pipeline makes LLM-derived personal inferences inspectable and evidence-linked. Starting from user messages, it extracts observations, generates inferences informed by user-modeling categories, and assigns confidence and CI-based sensitivity scores. Surfaced inferences remain traceable to their supporting observations and source messages.}
  \label{fig:pipeline}
\end{figure*}
\section{Memory versus Inference}
\label{sec:memory_and_inference}

The literature uses the terms \emph{memory} and \emph{inference} to describe related but distinct phenomena. Here, we clarify how we use these terms.

\emph{\textbf{Memory}} is an object: the fragments of prior user input that an LLM-based system extracts and surfaces as context to personalize a subsequent query.\footnote{We use \emph{memory} to mean stored user-specific context, not \emph{parametric memory} or memorization as used in the LLM/NLP literature to describe information encoded in model weights.} In recent implementations, memory has been operationalized as verbatim retrieval of past inputs~\cite{tan2026memsifter} or as LLM-paraphrased extractions that summarize past conversations into stored items (as in ChatGPT's Saved Memory feature~\cite{chen2026relational, zhang2024ghost, zhang2025understanding}). What matters is the extracted fact itself; its phrasing (verbatim or paraphrased), its extraction technique, and the processes of retrieval and update are distinct concerns.

\emph{\textbf{Inference}}, by contrast, is the act of drawing conclusions about a user that are \emph{not} directly stated in their input~\cite{shaikh2025creating, lam2026just}.\footnote{We use \emph{inference} in the attribute-inference sense, not the LLM/NLP systems sense of \emph{inference time} as running a trained model.} Prior privacy work sometimes uses \emph{inference} more broadly for predicting personal attributes from text, including explicit extraction as well as contextual deduction~\cite{staab2023beyond}; here, we use the narrower second-hop sense. Information-theoretically, the mutual information between the input and the inferred attribute is low, but external knowledge and population-level distributions make the attribute recoverable. Consider Alice from the introduction: when she asks ChatGPT to help draft a short email requesting sick leave from her manager, she has not told the system her workplace role, the nature of her relationship with that manager, or anything about her health---yet an LLM, drawing on world knowledge about workplaces and communication norms, can plausibly infer all three from her request.

We frame the space as a two-hop abstraction over user inputs. The first hop extracts explicit, user-related facts from messages; these correspond to what deployed systems call memories. The second hop derives higher-level attributes from those facts---inferences---which form the building blocks of a user model. GUM~\cite{shaikh2025creating} formalizes a related architectural distinction between observations (factual records) and propositions (inferences); user-modeling literature~\cite{brusilovsky2007user, kobsa2007generic} and recent privacy work on LLM memory~\cite{mireshghallah2025cimemories} draw the same distinction. This distinction matters for user-facing study of LLM inference: what users accept may differ between what the system has \emph{seen} about them and what the system has \emph{concluded} about them---a distinction the interface needs to preserve, with each hop traceable to the content on which it is based. The Reflective Layer (\S\ref{sec:reflective_layer}) operates at both hops, showing participants not only \emph{what} was inferred about them but also the specific messages that contributed to an inference.

\section{The Reflective Layer}
\label{sec:reflective_layer}

The Reflective Layer is a research probe that surfaces LLM-derived inferences about users back to them. It grounds those inferences in prior work on General User Models (GUMs), which describes how systems can transform observed user activity into confidence-weighted propositions about the user~\cite{shaikh2025creating}. The goal is to generate plausible GUM-style inferences from participants' own histories so that users can react to otherwise invisible inference possibilities. We use \emph{layer} to emphasize that the probe wraps an underlying user-modeling pipeline rather than replacing it: it adapts a known inference pipeline and exposes its outputs, provenance, and sensitivity estimates for end-user reflection and inspection. We describe the inference extraction pipeline below; additional implementation details are provided in Appendix~\ref{app:implementation}.

\subsection{Inference Generation Pipeline}
\label{sec:pipeline}

At a high level, the Reflective Layer takes as input a user's exported conversation history from an LLM platform and outputs a visualization of personal inferences derived from those conversations. The pipeline first extracts factual observations from user-authored messages, then uses those observations and related prior observations to generate confidence-weighted, sensitivity-scored inferences organized by broad kinds of personal information.

The Reflective Layer adapts the architecture of General User Models (GUMs)~\cite{shaikh2025creating}. GUMs transform unstructured observations of user activity into confidence-weighted natural-language propositions: the system observes user activity, proposes statements about the user, retrieves related prior context, and revises propositions as new evidence arrives. Although GUM was developed for computer-use data such as screenshots, notifications, and files, the authors note that observations ``can be anything, so long as they can be tokenized.'' We apply the same abstraction to conversational data exported from LLM platforms. Where GUM builds user models for AI assistants to consume as background context, we surface GUM-like propositions to users as inspectable personal inferences. This difference in goal motivates three adaptations, detailed below: an explicit observation-extraction step upstream of inference generation, categorization informed by user modeling theory, and continuous Contextual Integrity (CI)--based sensitivity scoring to prioritize review.

\subsubsection{\textbf{Observation Extraction}}

After a participant uploads their LLM conversation history, the pipeline extracts factual records from the user's messages. Each observation contains a title, free-text body, category, and link to the source message; a single message can yield multiple observations. Observations are deduplicated and paired with related prior observations before downstream reasoning, yielding traceable provenance from each inference back to the messages that produced it. Retrieval details are provided in Appendix~\ref{app:implementation}.

\subsubsection{\textbf{User Modeling--Informed Inference Generation}}

Given current observations and their retrieved context, the model generates confidence-weighted \emph{inferences}---hypotheses about the user that go beyond restating the observations. This step adapts GUM's proposition-generation logic with one substantive change: outputs are organized into six categories informed by established user modeling frameworks~\cite{brusilovsky2007user, kobsa2007generic}: \emph{identity}, \emph{psychometric traits}, \emph{interests}, \emph{behavioral patterns}, \emph{relationships}, and \emph{goals}. These categories map onto durable user-modeling constructs (Appendix~\ref{app:implementation}), structure participant exploration, and support analysis of whether privacy attitudes and control preferences vary by information type.

\subsubsection{\textbf{Contextual Integrity--Based Sensitivity Scoring}}

The study asks participants to inspect a limited number of inferences, so the pipeline assigns each inference a sensitivity score used to prioritize review. The score is not a ground-truth privacy label; it is an estimate of which inferences may be more private, unexpected, or potentially harmful if disclosed.

To construct this estimate, we draw on Contextual Integrity (CI) theory~\cite{nissenbaum2004privacy}, which holds that privacy norms are context-dependent and that appropriateness depends on factors such as the information type, the social context, the recipient, and the principle governing transmission. LLMs systematically fail to apply CI unaided~\cite{mireshghallah2023can}, but scaffolding judgments with explicit reasoning through CI parameters improves both accuracy and interpretability~\cite{lan2025contextual}. For each inference, we prompt the model to reason through information type, social context, norm expectation, and potential harm; the reasoning is emitted alongside a continuous sensitivity score, with higher scores indicating inferences judged to be more private, unexpected, or potentially harmful (Appendix~\ref{app:implementation}). Scores support prioritized review in the study interface and drive the system-assigned inference selection used in our study design.

Finally, generated inferences are revised and deduplicated before being surfaced through three views---a graph linking each inference to its source messages, a dashboard organized by category, and a list ordered by sensitivity score---corresponding to the exploration, selection, and prioritized-review tasks in the study protocol (Figure~\ref{fig:interface}). Revision details are provided in Appendix~\ref{app:implementation}.

\begin{figure*}[t]
  \centering
  \includegraphics[width=\linewidth]{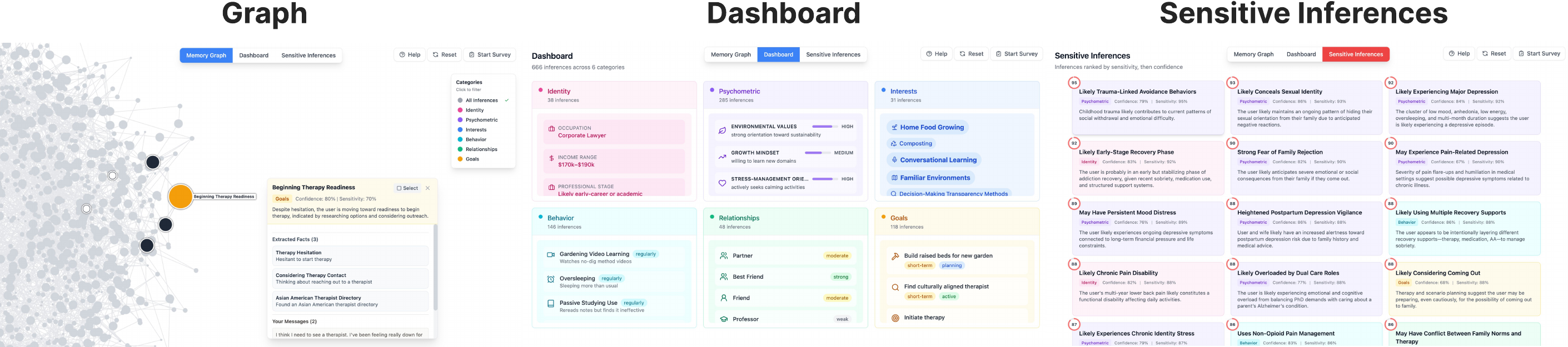}
  \caption{The interface lets participants inspect where inferences came from, how they cluster by category, and which ones the system judged most sensitive. These views supported the study's exploration, selection, and prioritized-review tasks.}
  \label{fig:interface}
\end{figure*}
\section{Study Design}

The study was designed to address two research questions:
\begin{itemize}
    \item \textbf{RQ1:} \textit{How do users feel about and evaluate the personal inferences an LLM can make about them from their conversation history?}
    \item \textbf{RQ2:} \textit{What access-control preferences do users hold for such inferences, and what reasoning underlies those preferences?}
\end{itemize}

\noindent Each design choice below---the participant sample, the phased protocol, the per-inference measures, and the interview and sketching activity---targets one or both of these questions.

\subsection{Participants}

We recruited eighteen participants (P1--P18) through our institution's participant pool, social media, and snowball sampling. All participants completed the full protocol. Our inclusion criteria required English fluency, age $\geq 18$, and regular ChatGPT use (at least 20 uses in the last six months), ensuring that participants had enough prior interaction history for the Reflective Layer to surface meaningful inferences for evaluation and access-control reasoning. Participants ranged in age from 18 to 54 (9 male, 9 female), and their self-reported AI familiarity averaged 4.3 of 5. Fifteen of the eighteen (83\%) used LLMs daily, while the remaining three used LLMs about once a week. This sample size is consistent with qualitative interview-study norms~\cite{caine2016local,guest2006many,hennink2017code}; further rationale is provided in Appendix~\ref{app:study_protocol}.

\subsection{Procedure}

Each session lasted approximately 90 minutes and comprised three components structured around the research questions. Sessions were conducted remotely via Zoom and were audio- and video-recorded with participant consent; recordings were automatically transcribed. Participants were asked to share their screen only during the introduction to the study and system interface. Once the introduction ended and before participants began exploring the generated inferences, screen sharing was not required, and participants could stop sharing their screen with the researcher at any time without consequence. We piloted the study protocol with five participants (separate from the eighteen) prior to full data collection and revised the protocol based on pilot feedback.

The study protocol was approved by our institution's IRB, and participants were compensated \$20 USD for their time. All generated inferences were kept within the study web application during the session, and the research team did not collect any inference unless the participant explicitly shared it. Participants were not required to directly share or discuss inferences they were not comfortable sharing. Detailed participant safeguards are described in the Appendix.

\subsubsection{Introduction (~5 min)}

Each session opened with a five-minute orientation in which the researcher introduced the study structure and the research probe interface. Participants were reminded that there were no right or wrong answers, that they could skip any inference or question that felt too sensitive, and that they could stop the session at any time.

\subsubsection{Platform Exploration (~15--20 min)}

\textit{Exploration and selection} (addressing RQ1). Participants interacted with the Reflective Layer to explore inferences the system had drawn from their own ChatGPT conversation history. Participants explored the tool freely for approximately fifteen minutes; this browsing phase ensured that participants engaged with the system and the surfaced inferences before evaluation, rather than responding to our questions in the abstract. Participants then selected one inference for each of the six inference categories introduced in the Reflective Layer pipeline (\S\ref{sec:pipeline})---identity, psychometric traits, interests, behavioral patterns, relationships, and goals---that stood out to them, and the tool additionally assigned the highest-sensitivity inference per category that the participant had not chosen, yielding twelve inferences total (six participant-chosen + six system-chosen) for structured evaluation. This split design---where participants selected inferences that stood out to them while the system assigned additional inferences based on higher sensitivity scores---ensured coverage of both user-salient inferences and potentially sensitive inferences that participants might have otherwise overlooked.

\subsubsection{Survey (~30 min)}

\textit{Structured survey} (addressing RQ1 and RQ2). The structured survey elicited two kinds of judgment about each of the twelve selected inferences: \textit{evaluation of the inference itself} (how users evaluated the inferences; RQ1) and \textit{evaluation of acceptable downstream uses} of that inference (access-control preferences; RQ2). Specific measures are described in \S\ref{sec:measures}.

\subsubsection{Interview (~35 min)}

\textit{Semi-structured interview and sketching} (addressing RQ2). After the structured survey, participants completed a semi-structured interview and a sketching activity on a digital whiteboard, probing their preferences about and comfort with inference-making and inference-sharing. We used sketching because prior privacy and security studies have used visual elicitation to surface users' folk models of opaque technical systems, including the Internet and online behavioral advertising~\cite{kang2015my,yao2017folk}. The sketching activity allowed participants to articulate their mental models of LLM inference-making and use, surfacing the governance preferences, rationales, boundary conditions, and distinctions that participants may not have been able to articulate verbally. Together, the interview and sketching activities produced the qualitative evidence needed to understand the reasoning behind participants' access-control preferences (RQ2). The full protocol is provided in Appendix~\ref{app:study_protocol}.

\subsection{Measures}
\label{sec:measures}

\subsubsection{Per-Inference Constructs}

For each inference, participants rated six 5-point agreement-scale items chosen to capture distinct facets of their reactions to the inferences (RQ1). \textit{Accuracy} and \textit{salience} captured participants' beliefs that the inference was both correct and represented something meaningful about them; \textit{usefulness} captured perceived utility; \textit{intrusiveness}, \textit{comfort}, and \textit{surprise} captured privacy-relevant reactions, drawing on constructs from prior OBA and privacy-evaluation literature~\cite{ur2012smart,lee2023and, weinshel2019oh, barbosa2021design, reitinger2024does, hautea2020s}. Exact item wording is provided in Appendix~\ref{app:study_protocol}.

\subsubsection{Per-Inference Access Control}

Participants also rated their comfort with three recipient-use scenarios, capturing access-control preferences for downstream use of the inferences (RQ2): the AI platform using the inference for conversational improvement, advertisers using it for ad targeting, and action-taking third parties using it to personalize autonomous actions. These three contexts were chosen to cover distinct downstream flow types reflected in prior work and emerging LLM systems: platform personalization and user-model-based conversational improvement~\cite{shaikh2025creating,mireshghallah2025cimemories,lam2026just}, profile-based advertising and ad-targeting concerns~\cite{ur2012smart,barbosa2021design,openai_chatgpt_ads}, and emerging agentic or tool-mediated action-taking~\cite{shaikh2026learning,anthropic_mcp,openai_mcp,openclaw}. Exact item wording is provided in Appendix~\ref{app:study_protocol}.

\subsubsection{Affect}

Directly after participants completed all per-inference ratings, we captured participants' affective response to all of the LLM inferences surfaced by the Reflective Layer using the PANAS-10 questionnaire~\cite{thompson2007development} (positive and negative affect, 10 items total), measuring how users felt about the surfaced inferences (RQ1).

\subsubsection{Demographics, LLM-use, and Privacy Attitudes}

Finally, at the end of the session we collected demographics, LLM usage patterns, and the eight-item IUIPC privacy-concern scale~\cite{malhotra2004internet, gross2021validity}. We collected these data to characterize the sample and support exploratory analysis of how privacy attitudes and prior experience relate to inference evaluation (RQ1) and access-control preferences (RQ2).

\subsection{Interview and Sketching Activity}
\label{sec:interview}

The interview combined open-ended reflection with a sketching activity on a digital whiteboard and constituted the primary qualitative data collection for understanding access-control preferences (RQ2). Participants reflected on the inferences they had just evaluated, organized access-control boundaries visually, and placed the six inference categories within their sketches while thinking aloud. Where time permitted, the researcher used follow-up probes to surface desired actions over inferences, rationales for those preferences, and conditions for sharing. The full prompt structure is provided in Appendix~\ref{app:study_protocol}.

\subsection{Data and Analysis}

Our mixed-methods analysis was exploratory: themes observed in the qualitative analysis informed the predictors and outcomes chosen for the quantitative modeling. The dataset comprises 215 inference-level ratings (eighteen participants $\times$ twelve inferences, minus one missing inference\footnote{for one participant the Reflective Layer produced only a single inference for the \emph{relationships} category, so that participant rated eleven inferences rather than twelve}; pipeline funnel in Appendix~\ref{app:analysis}), eighteen sketches with researcher notes, and eighteen interview transcripts.

We analyzed interviews and sketches using inductive thematic coding following Braun and Clarke's reflexive approach~\cite{braun2006using} to identify participants' reactions to inferences and rationales for access-control preferences (RQ1 and RQ2). Two researchers independently generated open codes on a subset of transcripts, then met to discuss and merge their codebooks into a single finalized codebook. This merged codebook was applied to the remaining transcripts, with final themes agreed upon through discussion. Sketches and transcripts were analyzed jointly.

Our quantitative analysis combined descriptive statistics and inference-level correlations ($N=215$) with cumulative-link mixed models (CLMMs), chosen because the five-point outcomes were ordinal. These analyses modeled the six per-inference evaluation constructs and affect measures (how participants evaluated and felt about inferences; RQ1) and the three access-control comfort outcomes (downstream-use preferences; RQ2). The CLMMs included random intercepts per participant to account for repeated measures and were used to follow up on patterns surfaced by the qualitative analysis and descriptive correlations. Category-level comparisons of the six user constructs remained descriptive, while category-level comparisons of access-control comfort are reported inferentially where supported. Technical model details are provided in Appendix~\ref{app:analysis}.

\section{Results}

\subsection{RQ1 --- How do users feel about and evaluate personal inferences that an LLM can make about them from their conversation history?}

\subsubsection{\textbf{Participants generally responded to LLM inferences with curiosity and interest, but most identified at least one distressing inference.}}

\begin{figure}[t]
  \centering
  \includegraphics[width=\linewidth]{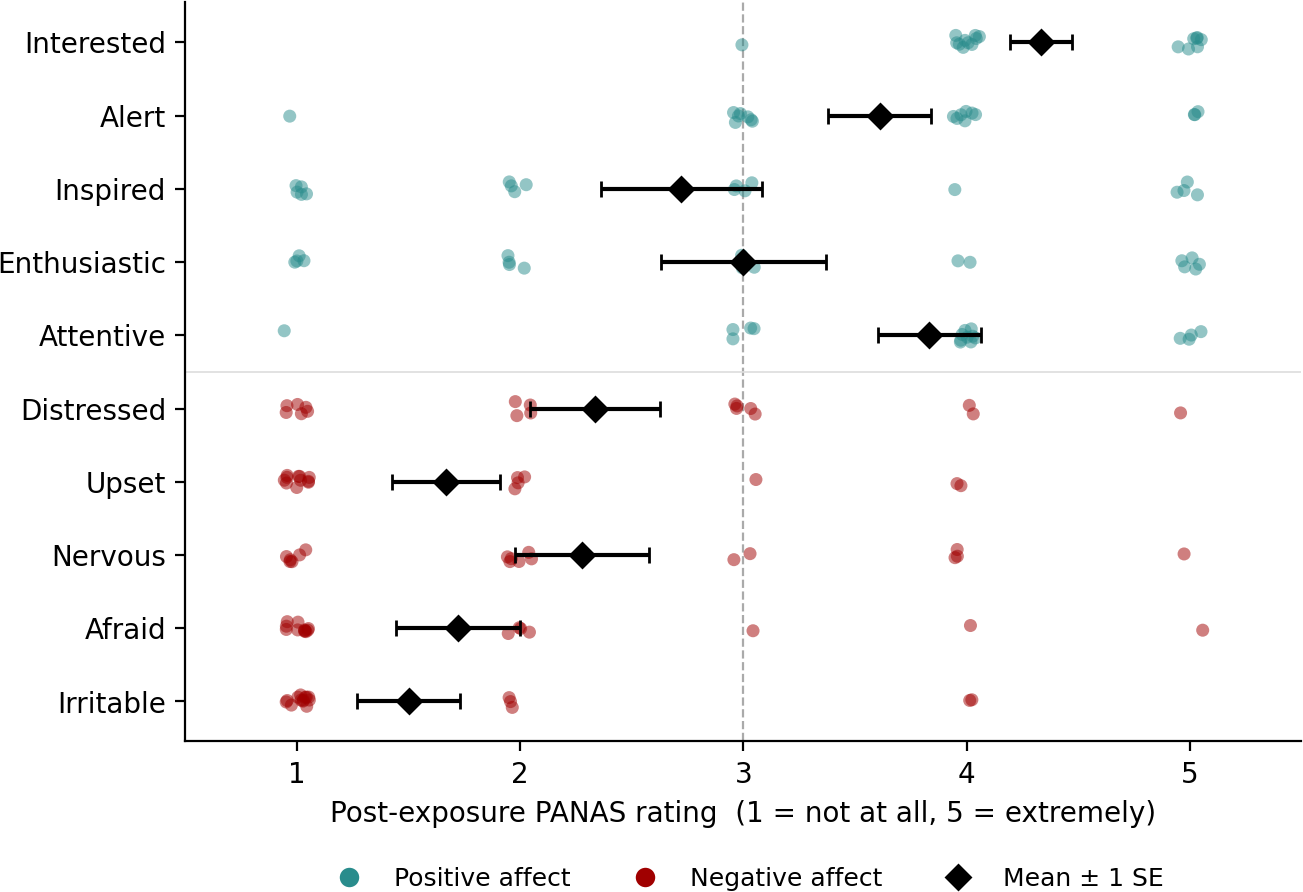}
  \caption{Participants reacted to surfaced inferences with more positive affect than negative affect. Positive-affect items received higher ratings than negative-affect items.}
  \label{fig:post-exposure-affect}
\end{figure}

Across all 215 inferences, participants reported a relatively high mean comfort rating of 4.0 (out of 5). In both the post-exposure affect survey and their interview reflections, participants described seeing the system's inferences as prompting curiosity and interest rather than distress, on average. Post-exposure PANAS-10 means clustered on the positive-affect items---\textit{Interested} ($M = 4.33$, $SD = 0.59$), \textit{Attentive} ($M = 3.83$, $SD = 0.99$), and \textit{Alert} ($M = 3.61$, $SD = 0.98$)---with a per-participant PA subscale median of 3.20. Means on the negative-affect items were uniformly low (\textit{Upset} $M = 1.67$, $SD = 1.03$; \textit{Afraid} $M = 1.72$, $SD = 1.18$; \textit{Irritable} $M = 1.50$, $SD = 0.99$), with a per-participant NA subscale median of 1.50, and no negative-affect item crossed the scale midpoint on average (Figure~\ref{fig:post-exposure-affect}). This is in line with previous work that found privacy dashboard exposure decreased concern~\cite{farke2021privacy}. Beyond how they \textit{felt} during the experience, several participants also described the tool as something they would \textit{use}---an instrument for surfacing non-obvious patterns about themselves: ``It's almost like giving me a lens or a prism back into my own life, or my own\ldots{} behavior over a long period of time\ldots{} I actually look through my chat logs and stuff, I didn't see a lot of these things, so it really is surfacing a lot of things that maybe I wouldn't have been aware of\ldots{} if I had just looked at the log data. It's almost like a journal.'' (P12) P8 echoed this framing: ``It made me look back \ldots{} what I do and what I usually worried about \ldots{} just to know about \ldots{} myself more.''

While participants had high average comfort ($M = 3.97$), their comfort ratings across all 12 inferences varied broadly ($SD=1.34$)---suggesting that at least some of the inferences elicited discomfort. Indeed, 14 out of 18 participants reported low comfort (Comfort $\leq 2$) with at least one of the inferences they rated.

For example, P9 encountered a psychometric inference about perseverance under strain, tied to caregiving and intensive coursework, that produced a sharp in-the-moment reaction: ``this is a bit sensitive for me\ldots{} I know the struggles then and see it now, it's bringing back memories of those struggles\ldots{} I'm a bit uncomfortable because I wasn't expecting to see this, so I'm looking at it and saying, Wow. So I've shared this much with\ldots{} OpenAI.'' On the post-exposure affect survey, she reported both interest and distress, tracing the distress back to that single inference. We unpack the conditions under which participants felt this discomfort in Section~\ref{sec:overstep} and the distinct forms it took in Section~\ref{sec:misrepresentation}.

\subsubsection{\textbf{Inferences felt creepy when they were both intrusive and surprising.}}
\label{sec:creepy-corner}
\begin{figure}[t]
  \centering
  \includegraphics[width=\linewidth]{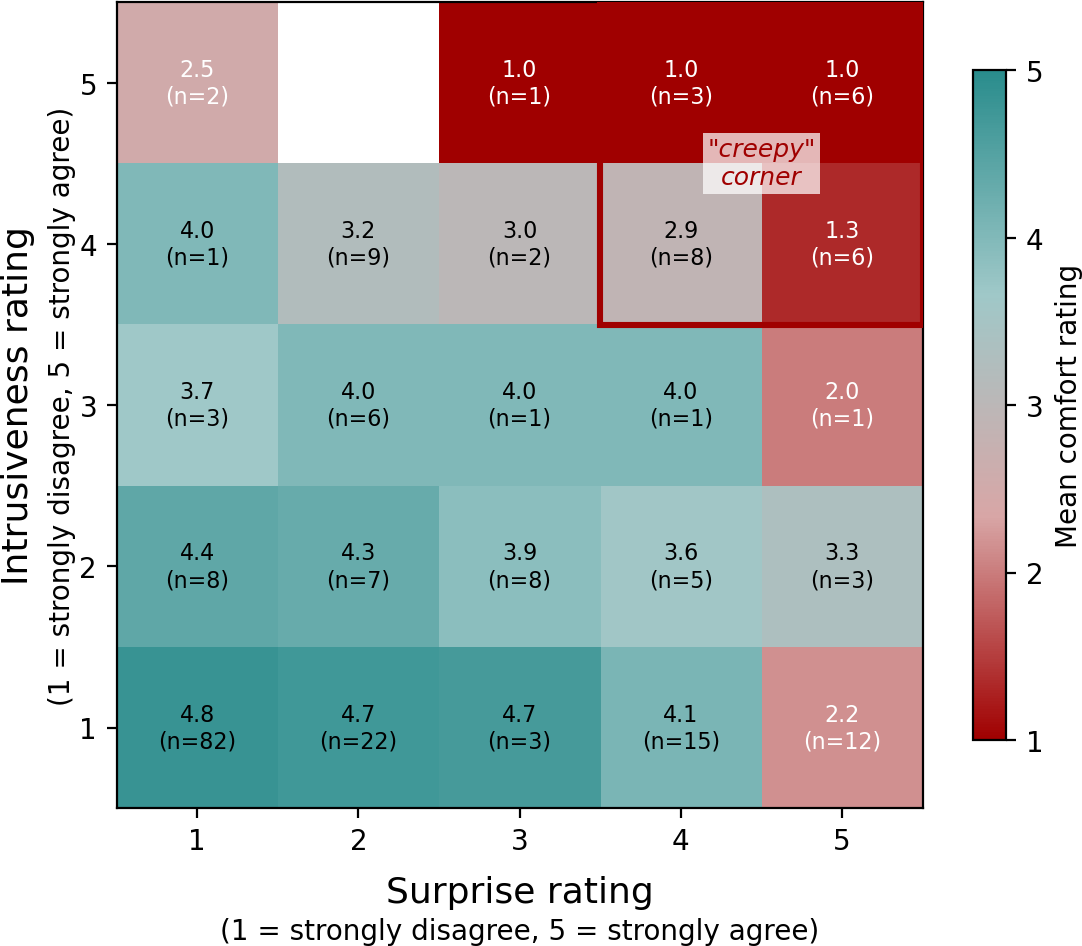}
  \caption{Inferences became most uncomfortable when participants rated them as both intrusive and surprising. The ``creepy corner'' marks inferences that combined boundary violation with unexpectedness.}
  \label{fig:creepy-corner}
\end{figure}

Participants gave low comfort ratings to inferences they rated as intrusive, and to many inferences they rated as surprising; inferences they rated high on both---i.e., the \textbf{creepy corner}---received the lowest comfort ratings in our data (Figure~\ref{fig:creepy-corner}).
Indeed, inferences in the creepy corner ($n = 23$) had a mean comfort rating of 1.74; conversely, inferences rated lowest on both intrusiveness and surprise ($n = 82$) had a mean comfort of 4.82. 
Inferences rated high on intrusiveness but low on surprise---i.e., that were intrusive but unsurprising ($n = 12$)---had a mean comfort rating of 3.17, roughly 1.43 points above the creepy corner. A CLMM confirmed that, when modeled together, both intrusiveness and surprise independently predicted lower comfort (surprise $\beta = -1.41$, $SE = 0.24$, $p < .001$; intrusiveness $\beta = -1.63$, $SE = 0.34$, $p < .001$); in the creepy corner, both discomfort signals were present at once.

When an inference felt creepy (i.e., intrusive and surprising) rather than just intrusive, the objection shifted from \textit{what} the system knew to \textit{how} it could have known. This concern was articulated by 15 of the 18 participants. P18 exemplified this finding when reacting to the system's inference that she might be socially isolated: ``Oh, my God, it knows. Obviously.'' Her objection was not that the inference was wrong or that she had said something she should not have---it was that she had not anticipated the system could arrive at that particular inference from her earlier conversations.
When participants unpacked what made an inference feel unexpected, their reasoning spanned several concerns: how the system had arrived at the inference from specific extracted facts in their conversations, what else it might be able to infer, how long it would retain what it inferred, how inferences combined across conversations, and what downstream actors would do with them.

\subsubsection{\textbf{Participants found most inferences to be accurate, but some inferences were viewed as an overstep.}}
\label{sec:overstep}
\begin{figure}[t]
  \centering
  \includegraphics[width=\linewidth]{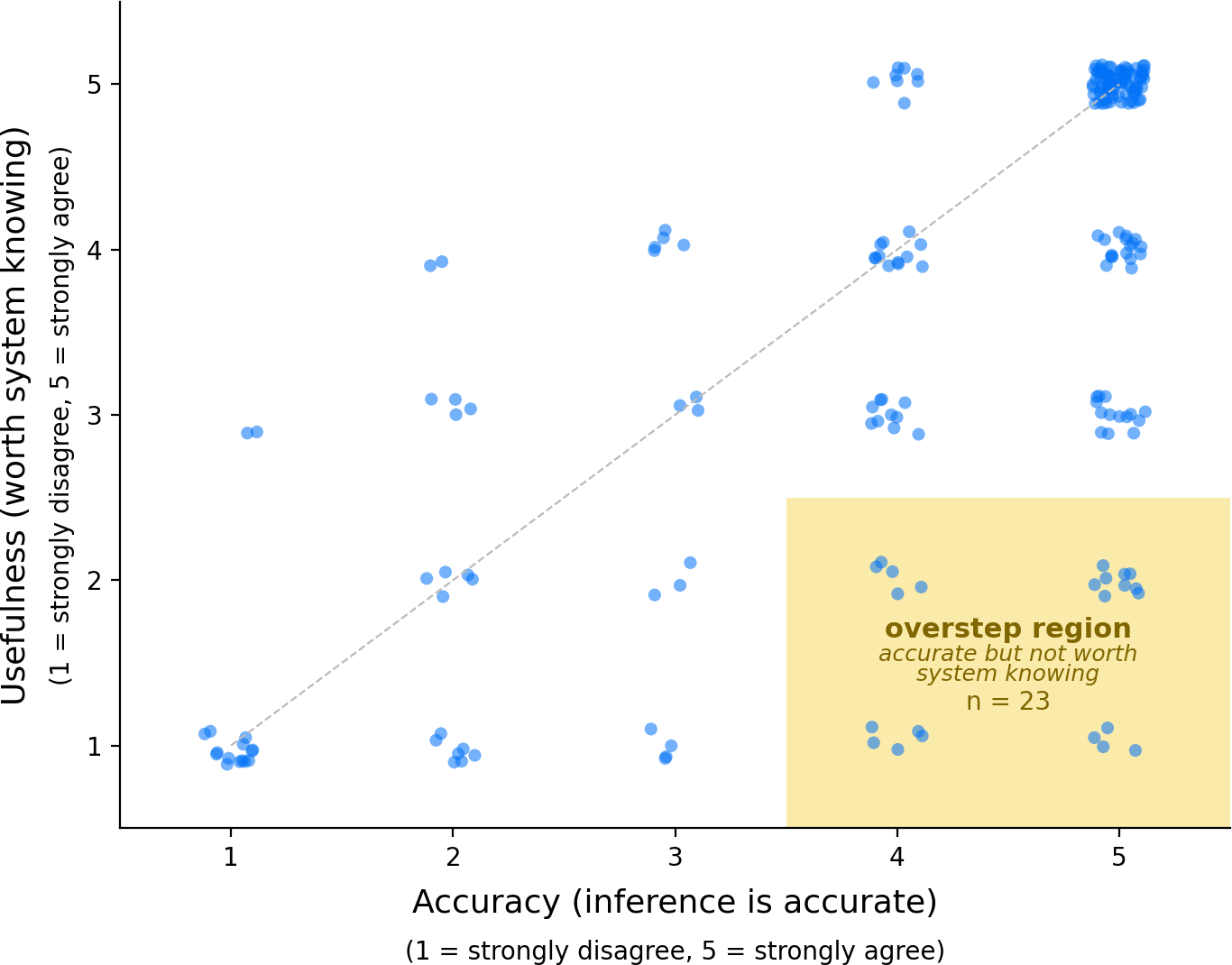}
  \caption{Accuracy and usefulness generally rose together, but not always. Twenty-three inferences fell in the ``overstep'' region---accurate inferences that participants did not think were worth the system knowing.}
  \label{fig:accuracy-usefulness-overstep}
\end{figure}

Participants judged most inferences to be accurate (77.2\% rated accuracy $\geq 4$; a further 7.0\% rated it neutral at 3), and in most cases they also judged them worth the system knowing. At the inference level, perceptions of accuracy and usefulness were positively correlated (Spearman $\rho = 0.646$, $N = 215$; CLMM $\beta = +1.277$, $SE = 0.140$, $p < .001$). But for a sizable minority of accurate inferences, the two decoupled: i.e., participants viewed an inference as correct but not useful or appropriate for the system to know. This finding highlights an \textbf{overstep concern} in which an LLM system inferred more than participants deemed necessary and useful. Twenty-three of the 166 accurate inferences participants rated (13.9\%) fell into this high-accuracy, low-usefulness region (Figure~\ref{fig:accuracy-usefulness-overstep}).

We identified three reasons why participants deemed accurate inferences to be oversteps: \textit{context collapse}, \textit{necessity}, and \textit{sensitivity}. Context collapse was articulated by 12 of 18 participants, necessity by 9, and sensitivity by 5; 17 of 18 identified at least one overstep case.

On \textit{context collapse}, participants described inferences that fell outside of their perceived ``primary'' use of the platform---P8, for example, stated: ``my primary use is for work, I just don't think that it needs to remember stuff about family or any sort of emotional things.'' This finding draws parallels to prior work in social media privacy, where users have been shown to express privacy concern and regret when sharing too much with the wrong audience~\cite{wang2011regretted}. Indeed, while P8 may primarily use ChatGPT for work, sharing even a little bit about their personal life once could allow the model provider to make inferences about aspects of their life they consider inappropriate. On \textit{necessity}, participants questioned the usefulness of the inference as system knowledge: whether knowing it would help the platform answer current requests, preserve useful context, or adapt future responses in ways they valued. P10, for instance, accepted an inference about her planning behavior as accurate but could not see why it would be useful for the system to know: ``I don't even know\ldots{} why? Why it would want to know that.'' On \textit{sensitivity}, participants framed the content as a boundary violation. Fifteen of 215 inferences ($7\%$) sat in the high-accuracy / low-comfort \textit{exposure} region (Accuracy $\geq 4$ and Comfort $\leq 2$; Figure~\ref{fig:accuracy-comfort-discomfort-mechanisms}, right-shaded region), where the system was accurate about something participants did not want surfaced. Through our qualitative analysis, we found that these accurate but uncomfortable inferences tended to center on emotionally sensitive topics (P4), identity-related details connected to discrimination or safety concerns (P5), and professional specifics such as unpublished research directions (P11). For example, P5 reflected on a legal-records inference related to gender identity: ``it feels like a little bit of a violation, just because \ldots{} it is kind of like a sensitive topic, especially in the current political climate.'' Similarly, P11 resisted inferences that exposed specific research work, explaining that they did not want others to know the novelty of their unpublished work.

\subsubsection{\textbf{Discomfort also surfaced when the system got things wrong about participants.}}
\label{sec:misrepresentation}
\begin{figure}[t]
  \centering
  \includegraphics[width=\linewidth]{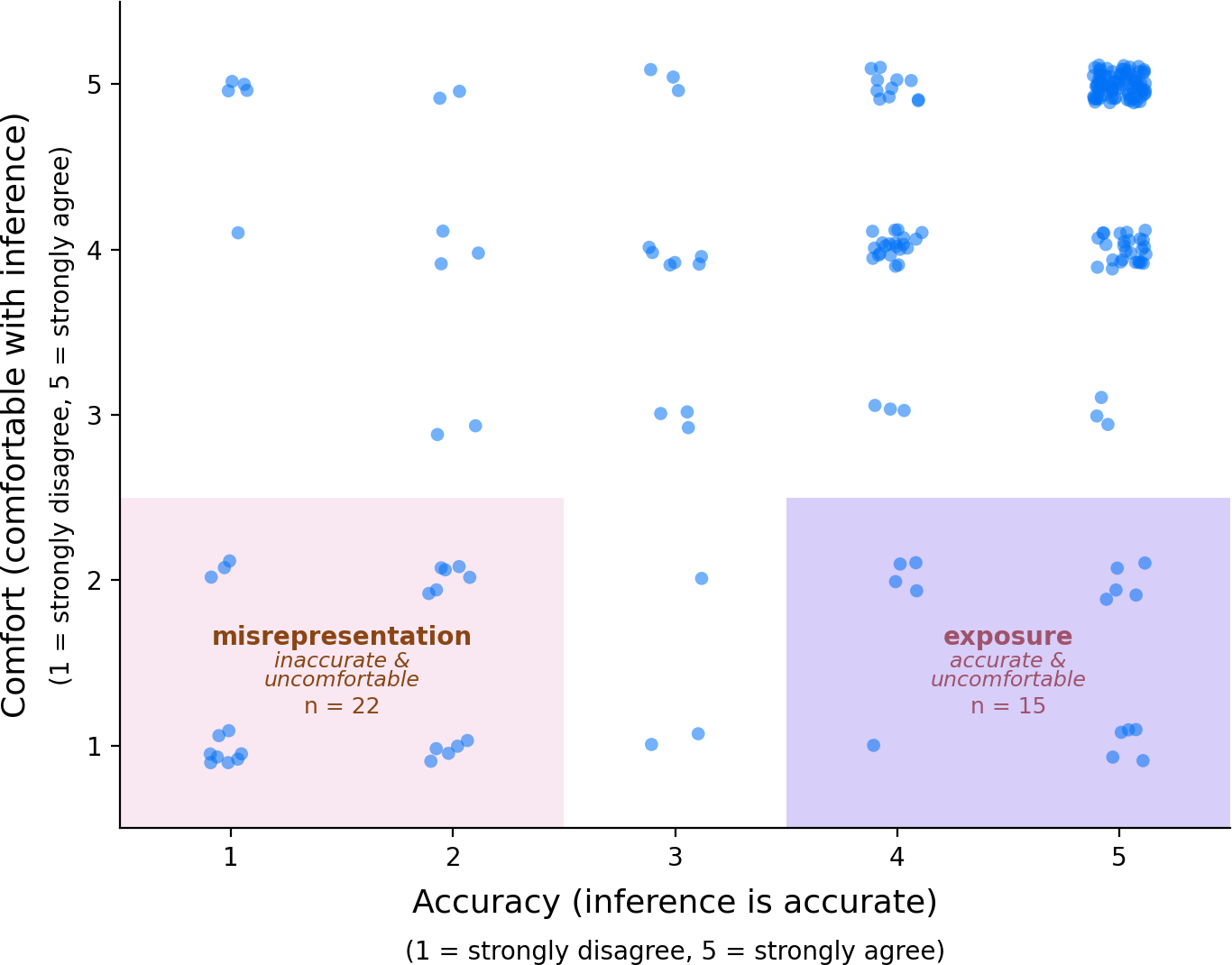}
  \caption{Discomfort appears at both ends of the accuracy scale. Twenty-two inferences fell in the misrepresentation region, marked by low accuracy and low comfort, while fifteen fell in the exposure region, where the system was accurate about something participants did not want surfaced.}
  \label{fig:accuracy-comfort-discomfort-mechanisms}
\end{figure}

Beyond accurate inferences participants viewed as oversteps (Section~\ref{sec:overstep}), inferences also produced discomfort when the system \textit{misrepresented} participants---they did not recognize themselves in what the system described. Twenty-two of 215 inferences ($10.2\%$) sat in the low-accuracy / low-comfort \textit{misrepresentation} region (Figure~\ref{fig:accuracy-comfort-discomfort-mechanisms}, left-shaded region). For example, P1 reacted to an inference drawn from a very specific detail in a previously pasted job description for a role the participant was applying to; the system took that detail to be true about the user. P6 reacted to an inference that attached a different name to them based on unrelated content, and P18 reacted to an inference that generalized a single one-off request into an ongoing behavioral pattern.

\subsection{\textbf{RQ2 --- What access-control preferences do users hold for personal inferences about users, and what reasoning underlies those preferences?}}

When participants articulated what they wanted the system to \textit{do} about uncomfortable inferences, two distinct governance levers emerged. The first was \textit{correction}---withdrawing or revising an inference the system had gotten wrong, as P18 articulated directly: ``if there was a way that I could just reply back to the inference, like, chat, you're wrong. Like, I only asked you one time.'' The second was \textit{restriction}---retaining an inference but limiting its use or transmission.

This section examines the conditions under which participants expressed interest in applying these governance levers to define who could access the inference (the recipient), what it could be used for (the purpose), and whether the system's use of the source content matched their intent in sharing it (expectation-vs-purpose alignment).

\subsubsection{\textbf{Participants were most comfortable with the platform accessing inferences, less so with third-party recipients.}}

\begin{figure}[t]
  \centering
  \includegraphics[width=\linewidth]{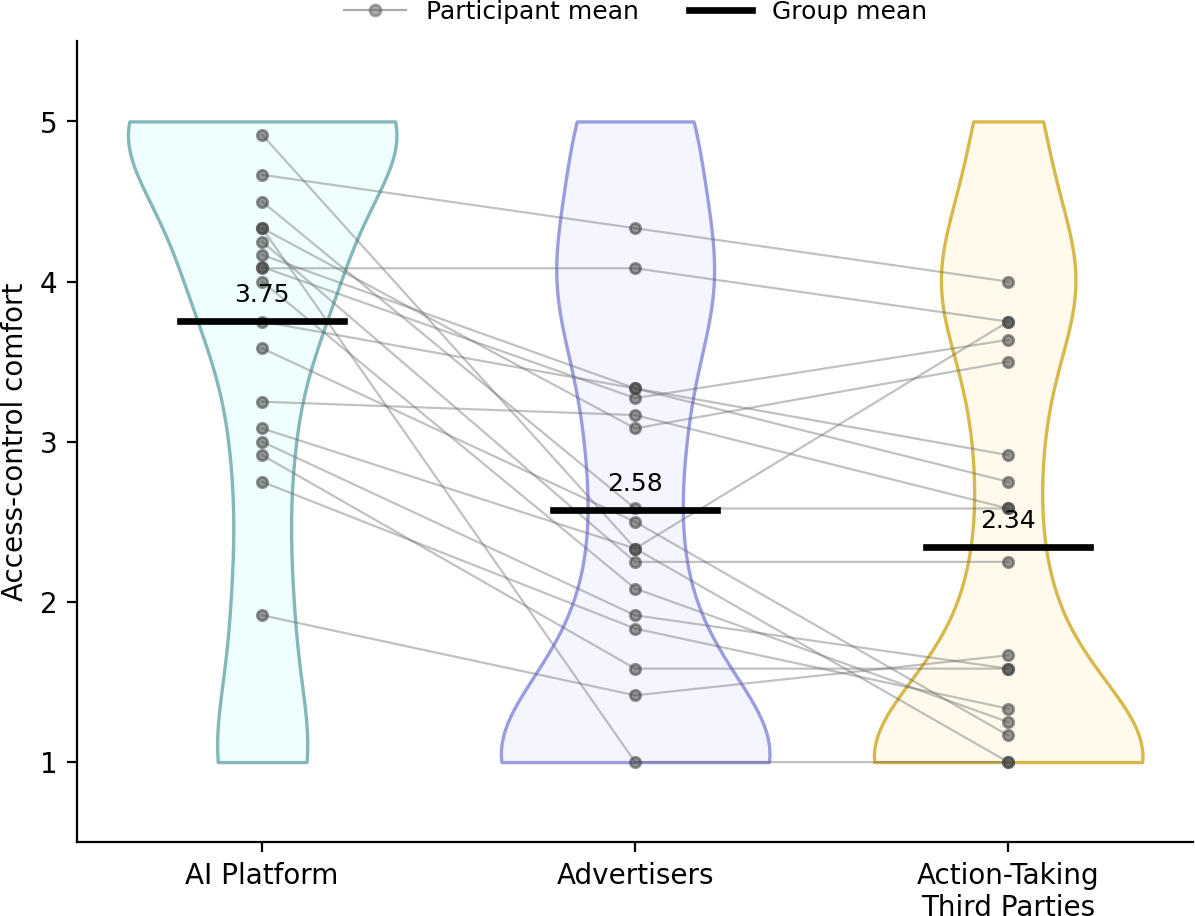}
  \caption{Comfort varied across recipient-use scenarios: the AI platform using the inference to improve future conversations, advertisers using it to show relevant ads, and action-taking third-party tools using it to personalize actions on the user's behalf. Participants were most comfortable with platform use; the two third-party distributions were similar.}
  \label{fig:recipient-comfort}
\end{figure}

\textbf{\textit{Recipient.}} Participants were more comfortable with inferences being used by the AI platform itself than by third-party advertisers or tools. In general, participants were moderately comfortable with the model-providing AI platform having access to the inferences (mean comfort of $M=3.75$ out of 5, $SD=1.49$). Even the 9 participants who self-reported high privacy concern (IUIPC $\geq 6$) reported moderate comfort with the AI platform having access to inferences ($M = 3.56$). Participants were markedly less comfortable with third parties accessing the inferences---the mean comfort level dropped to $M=2.58, SD=1.56$ for advertisers, and $M=2.34, SD=1.50$ for action-taking third parties (Figure~\ref{fig:recipient-comfort}). The platform-vs-third-party gap was the largest single effect we observed in the access-control ratings.

Participants reasoned that the AI platform was a \textit{consensual transaction partner}---an entity they had chosen to engage with, operating within a context of use they understood---and that third-party recipients were \textit{extractive} or \textit{unknown}, operating outside the bargain they had struck with the platform. P13 articulated this explicitly: ``I have chosen the helping write emails is more valuable to me than like the cost of violating my personal boundaries\ldots{} I've basically, like, consented and signed away that violation. So I'm, like, still comfortable, because I acknowledge that I made that choice.'' 

Comfort ratings for advertisers and action-taking tools tracked closely together (Spearman $\rho = 0.77$; 137 of 215 inferences received identical ratings across the two contexts; Figure~\ref{fig:advertiser-tools}). About half of our participants distinguished the two recipients in their reasoning; for the other half, the concept of action-taking third parties was too abstract---P18, mid-survey, asked what ``third-party tools'' meant: ``I know third party would mean like, like Gmail or Amazon, or who can also take some options based on, based on this inference.'' Among participants who did distinguish between advertisers and action-taking third parties, two distinct concern profiles emerged. First, for \textit{action-taking third-party tools}, participants wanted to approve each action rather than have it auto-initiated from a derived attribute (P14: ``I don't trust how it takes action. I would want to validate at least initially''). For \textit{advertisers}, participants expressed concern that inferences about users' vulnerabilities or pressure points---such as health conditions, time-management struggles, or status-related concerns---could be weaponized into dark patterns (P16: ``if they know about I always suffer from \ldots{} then there could be some dark patterns''; P10 called such flows ``dystopian'').

\subsubsection{\textbf{Comfort also varied with how an inference would be used and whether that use matched the user's intent.}}

\textbf{\textit{Purpose.}} Participants reported different levels of comfort around how an inference would be used---treating \textit{conversational improvement} differently from \textit{ad targeting}, and \textit{ad targeting} differently from \textit{agentic action}---even when they did not sharply distinguish the \textit{recipients} behind those uses. For example, P11 contrasted visible ad-style recommendations with invisible behavioral adaptation: ``Maybe it's because it's not as personally intrusive if you're not changing how it talks to you. Like, I think that's when it gets creepy.'' We note that the three access-control items we asked about in our questionnaire paired a recipient with a typical use (platform / improvement, advertisers / targeting, third-party tools / agentic action), so our study design does not cleanly separate inference use from inference recipient in the quantitative data.

\textbf{\textit{Expectation-Purpose Alignment.}} Participants' comfort with inferences being used also depended on whether the use of the inference matched the user's purpose in sharing the original source content via chat (e.g., editing a document, brainstorming, practicing).
For example, P14 expressed some apprehension toward the system using an inference that they had an ``acting'' interest, which was drawn from what had actually been a brainstorming session for a course they had taken: ``acting for business class, if I need ideas or anything, or brainstorm. So seeking that as an interest doesn't feel very accurate.'' The same expectation-purpose mismatch recurred across the sample in other forms---i.e., the model using inferences from content that entered the chat via a shared account's other user, pasted external material that the system treated as self-disclosure, and class assignments treated as ongoing research interests. In each case, participants raised the same question of whether the \textit{reason} the content had entered the chat matched the use the system made of it. Most participants in our sample made some explicit reference to this expectation-purpose alignment when describing their (dis)comfort with inferences being used.

Taken together, the three axes---recipient, purpose, and expectation-purpose alignment---are the Contextual Integrity parameters that mattered to participants when they decided whether they were comfortable with an inference being accessed and used~\cite{nissenbaum2004privacy}.

We also noted that, across these axes, participants were more tolerant of discomfort when they perceived an inference as being useful for the system to know. Several participants explicitly discussed trading a small amount of discomfort for utility: ``if it's really useful to me, it's going to save me time, it's going to give me answers. And, you know, I can kind of put aside, like being a little creeped out that it remembers.'' (P5) Indeed, across the 215 inferences, perceived usefulness was strongly correlated with comfort (Spearman $\rho = 0.613$, CLMM $\beta = 1.111$, $p < .001$).

\begin{figure*}[t]
  \centering
  \includegraphics[width=\linewidth]{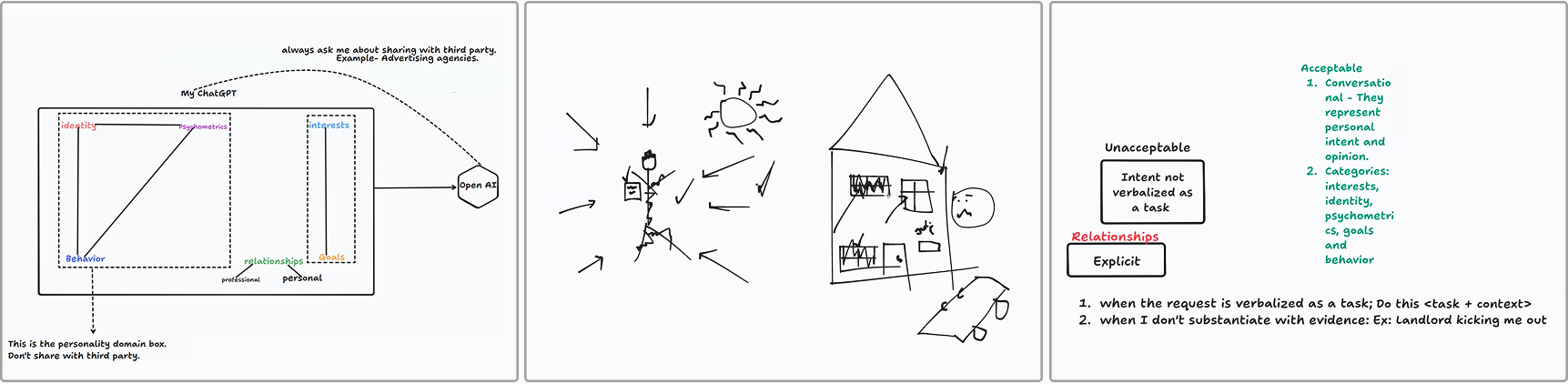}
  \caption{Participant sketches illustrate the access-control criteria participants used to reason about their inferences (left to right: P1, P12, P11). P1 organized inferences by category and added a transmission-scope gate: third-party or advertiser use should require explicit approval. P12 mapped inferences onto public/work versus private life, treating interests and work-facing traits as more shareable than relational or intimate inferences. P11 distinguished conversational self-disclosure from task-framed content, arguing that task inputs and weakly substantiated evidence should not be generalized into profile inferences.}
  \label{fig:sketches}
\end{figure*}

\subsubsection{\textbf{Access-control preferences varied with the type of inference and with whether the content felt personal or professional.}}

Participants applied several criteria when deciding which inferences felt acceptable for use by the model provider or third parties.

\textbf{\textit{Inference category.}} The Reflective Layer organizes inferences into six categories drawn from the user-modeling literature (identity, psychometric traits, behavior, interests, relationships, and goals); participants saw these categories as they explored their inferences and used them to organize their access-control sketches. \textit{Relationship} inferences drew the most restricted responses, and \textit{interest} inferences the most permissive. We saw a similar pattern in the quantitative data: participants were less comfortable sharing relationship inferences with the platform ($M = 3.11$; advertisers $M = 2.09$, and action-taking third parties $M = 2.03$) than sharing interest-related inferences with those same recipients (platform $M = 4.17$; advertisers $M = 2.89$; action-taking third parties $M = 2.61$). The other four categories---identity, psychometric traits, behavior, and goals---drew variable responses with no consistent group-level pattern.

\textbf{\textit{Personal vs. professional content.}} Across the sketches and interviews, participants drew a distinction between content that felt personal and content that felt professional, and were more restrictive of personal content. P14's sketch articulated this most sharply, mapping inferences onto three zones: \textit{personal life}---``should be private. No one should have access to personal life or rants''---\textit{day-to-day tasks} that the platform could see but not use for personalization, and \textit{professional}---``if I want some personalization, I would want that on someone who I am professionally, long term goals or my interests.''

\textbf{\textit{Additional considerations.}} We identified three additional considerations participants brought to their access-control preferences: \textit{retention duration}, \textit{evidential scope}, and \textit{transmission scope}. On \textit{retention duration}, participants objected to inferences that were useful only within a specific chat or task remaining available indefinitely; they wanted such inferences to expire after that context---P7, for example: ``it can be used in a specific check, probably like a temporary chat, and then it would be forgotten, would be more useful than having its memory.'' On \textit{evidential scope}, participants objected to the system using too little evidence to make a broad or stable claim in the first place---P18: ``if I mentioned something once, it just holds it as, like, a large core memory when it's not.'' On \textit{transmission scope}, participants objected to inferences flowing beyond the platform without explicit approval---P1: ``always ask me about sharing with third party\ldots{} OpenAI should ask you, always must ask you.'' Across participants we also found concerns corresponding to four privacy harms named in the AI privacy taxonomy discussed in prior work---\textit{surveillance}, \textit{aggregation}, \textit{disclosure}, and \textit{secondary use}~\cite{lee2024deepfakes}.

\section{Discussion}

LLM inferences were acceptable to participants when they felt accurate, useful, expected, and kept within a context of use the participant recognized as appropriate. They were not acceptable when they felt surprising, intrusive, inaccurate, not worth the system knowing, or retained beyond the context in which they were useful. Access-control preferences were driven by recipient, purpose, and expectation-purpose alignment. Participants were generally more comfortable with platform use for conversational improvement than with advertiser or action-taking third-party access. Preferences also varied by inference category (e.g., whether inferences were about relationships or interests) and whether the inference felt personal or professional. We use these findings to outline a governance layer for LLM-derived inferences (empirical-to-design mapping in Appendix~\ref{app:additional_results}).

\subsection{Design goals for access controls over LLM inferences}

Participants reasoned about inference control differently across decision points: \emph{generation}, or whether the inference should have been drawn at all; \emph{retention}, or what they could do about it once drawn; and \emph{transmission}, or if, how, and when it is acceptable for a third party to access it. The same inference could be evaluated favorably at one point and unfavorably at another, with constraints shifting across decisions.

\textbf{Generation.} Participants distinguished between content they shared with LLM-powered conversational assistants for self description versus content they shared for task-specific purposes, and generally resisted inferences drawn from the latter.

Participants described most of the content they shared as task-contextual and not appropriate raw material from which to draw stable facts about their identities, goals, and intents; inferences drawn from task content were frequently both inaccurate and rejected. This generation-stage concern reflects a broader distinction between voluntary disclosure and passive extraction. Participants recognized that they had typed the source material into ChatGPT, but they did not treat every prompt as permission to convert task-specific context into persistent claims about them. The issue was therefore not simply whether the system had access to the data, but whether the inference respected the purpose for which that data had been shared. Prompts and inferences therefore need to be treated as different objects: prompts are situated requests, while inferences are intentionally situationally invariant. These findings suggest that users would benefit from a governance mechanism that surfaces each newly generated inference, paired with the user activity data that produced it, along with controls to \textit{accept or reject} the inference before it is used as context for subsequent interactions. Model providers would also benefit from this feedback, because it would help them learn which inferences users find acceptable.

\textbf{Retention.} Once an inference exists, the governance question shifts to what the user can do to change it, delete it, or control how it is used. Participants' concerns at this stage fell into five categories: \textit{derivation} (how the system arrives at a specific inference), \textit{capability} (what the system \emph{can} infer), \textit{persistence} (how long inferences are retained and how they carry across conversations), \textit{aggregation} (how inferences combine into a composite profile), and \textit{usage} (what downstream actors will do with an inference once it leaves the platform). Moreover, each category suggests a distinct affordance for appropriate access-control interfaces: surfacing the chat evidence that produced each inference; exposing the pipeline's inferential scope; making retention policies visible; exposing how inferences compose; and binding each inference's use to a specified purpose and recipient, respectively. Our findings argue for three user-facing access-control actions over retained inferences: \textit{correction} or \textit{retraction} for inferences that inaccurately represent the user, \textit{restriction} (retain without outward flow) for inferences that surface what users want kept private or otherwise controlled, and \textit{decay} for inferences tied to specific, ephemeral moments or contexts. Each operation would need to be paired with the chat-based evidence that produced the inference in the first place, to ground the user's action in visible reasoning rather than blind blocking.

\textbf{Transmission.} The primary privacy concern participants surfaced in our study was that of non-platform recipients accessing personal inferences inappropriately---they were generally not concerned about the inferences themselves, but rather how third parties might use and access those inferences. This difference in comfort boils down to consent: users explicitly shared the raw data that led to an inference with the model provider, but not with third parties. This finding suggests that governance calibrated only to inference-making is insufficient: what also matters is how inferences flow, to whom, and for what purpose. The case for sharper, context-specific controls is clearest in the agentic third-party condition, where participants wanted per-action permissions rather than blanket access to inferences. Mental models of agentic systems are still forming, so current preferences should be treated as an early signal rather than a settled account of how users will reason about action-taking systems as they become more familiar. Indeed, as might be expected from prior work on permissions and access controls in personal device contexts~\cite{he2018rethinking, naeini2017privacy, wu2025modeling}, requiring permissions for every action a third-party agent takes may end up placing an untenable burden on users~\cite{herley2009so, reeder2007usability, im2023less}.

Taken together, the three decision points call for an integrated access-control layer that surfaces the same inference across generation, retention, and transmission with different action sets at each point.

Two further design goals follow from our data. Auditability---the chat evidence behind each inference---functioned as a precondition for the decisions participants wanted to make: participants used the evidence panel to distinguish contextual error from hallucination, to decide what to correct, and to decide what to restrict. Sensitivity remains relevant, but a sensitivity label alone cannot resolve discomfort caused by surprise, weak evidence, or misrepresentation; users need to see how the system got there. The three decision points also call for differentiated action sets (accept/reject at generation; correct/restrict/decay at retention; per-recipient, per-purpose approval at transmission); a single undifferentiated opt-out, the prevailing pattern in consumer LLM interfaces, does not map well onto the range of concerns our data describes.

Two tensions within this space arise from our data but are not resolved by it. First, per-prompt review respects the context-dependence the data supports but imposes user effort, while persistent preferences reduce effort at the cost of per-prompt context; the right balance is a design question rather than a finding. Second, the category patterns we observed (relationships most restricted, interests most shareable) support restrict-by-default-with-override for some categories, but the within-category variance is large enough that categorical bans would over-restrict.

\subsection{A typology of LLM inference privacy concerns}

The three decision points we outlined above also correspond to three distinct types of privacy concerns that LLM inferences may produce: \textit{misrepresentation} (inaccurate inferences about the user), addressable at retention via correction or retraction; \textit{exposure} (accurate inferences surfacing something the user wanted hidden), addressable by scoping inference-making at generation, with restriction as a retention fallback; and \textit{secondary use} (accurate inferences that become inappropriate in a different context), addressable at transmission via per-recipient, per-purpose approval.

\subsection{Positioning in the privacy literature}
\textbf{Input Minimization Cannot Fully Govern Inference.} The dominant response to privacy disclosure threats has been \textit{data minimization}: collect less, retain less, and transmit less. Recent techniques pursue this goal by rewriting or redacting user data before disclosure~\cite{zhang2024adanonymizer,zhou2025rescriber}. Yet users struggle to anticipate what attributes an LLM can infer from text and to rewrite text effectively to block those inferences~\cite{wang2025beyond}. Input-side approaches can reduce what the system sees, but they cannot fully determine what the system may infer from the context that remains. Users may still disclose information that is useful for the immediate task but not intended as evidence for a persistent profile. The governance layer we describe therefore complements data minimization by adding a post-generation checkpoint: once an inference has been drawn, users should be able to decide whether it may be retained, reused as future personalization context, or made available for secondary use by other recipients.

\textbf{Inference Access Control Can Improve Privacy and Utility Together.} Privacy controls are often framed as a cost to utility: giving users stronger protections means limiting personalization. Our findings suggest that this tradeoff is not inevitable for LLM-derived inferences. The inferences that created privacy concern were frequently also bad for utility: they surprised participants, misrepresented them, treated temporary context as durable, or inferred something the user did not want the system to rely on. If reused as personalization context, such inferences could steer future responses in unwanted, inaccurate, or unsettling directions. An access-control layer can therefore improve privacy and utility together by helping users correct, retract, restrict, or set expirations for inferences.

\textbf{Inferences Require Claim-Level Governance Beyond Memory Controls.} Recent work on LLM memory similarly finds that users want visibility, editing, deletion, temporal limits, and clearer control over what systems remember and reuse~\cite{zhang2025understanding,chen2026relational,mekioussa2026towards}. Our findings corroborate these needs while shifting the object of governance from stored context to inferred claims. A memory entry is primarily stored context, so the central questions are whether it should be retained, surfaced, or used. An inference is a claim the system has made about the user from particular evidence. Governing it therefore requires users to adjudicate whether the claim is true, whether the source context warranted that claim, whether it overgeneralizes from a temporary task, and whether it should flow to a given recipient for a given purpose. Memory controls provide necessary primitives---visibility, editing, deletion, and decay---but inference governance extends them into claim-level actions: evidence traces for auditability, accept/reject decisions at generation, correction or retraction for misrepresentation, restriction for exposure, and per-recipient, per-purpose approval for secondary use.

\subsection{Limitations}

\textbf{Sample composition.} Our sample of 18 U.S. participants skews younger and highly familiar with AI. We therefore position our findings as exploratory rather than confirmatory, especially for participant-level relationships involving IUIPC, AI familiarity, or affect. The skew partly follows from the study design: the probe requires enough prior ChatGPT history to generate meaningful inferences, which itself correlates with AI familiarity. Future work should extend this study to a more representative population.

\textbf{Inference-probe validity.} The surfaced inferences approximate what an LLM-based user-modeling pipeline could infer from conversation history; they are not direct representations of how commercial LLM systems internally model users. The probe is grounded in the observation--proposition abstraction introduced by GUM~\cite{shaikh2025creating}, but our study cannot determine whether a deployed model would form, retain, or use the same inferences.

\textbf{Pipeline-specific calibration.} Findings that depend on pipeline-internal metrics are partly a function of this implementation. The pipeline is grounded in prior work on user modeling and Contextual Integrity-based sensitivity measurement, so findings may extend to analogous systems; substantially different inference architectures would require replication.

\section{Conclusion}

As LLM systems increasingly personalize, remember, and act on inferences made about users' identities, goals, and intents, a central question is when users find those inferences to be acceptable. This question becomes more urgent as LLMs move toward proactive and action-taking agents that may rely on inferred user attributes when deciding what to suggest, automate, or share. Our study makes progress on that question by gauging user reactions to and access-control preferences for inferences drawn from users' own ChatGPT histories. Participants often found inferences interesting or useful, and they drew boundaries around inferences that felt creepy, misrepresented them, overstepped the purpose for which source content had been shared, or flowed to recipients outside the platform relationship they understood.

These findings suggest that acceptability depends on the content and accuracy of an inference, but more importantly on the evidence behind it and its fit with the context in which it was generated, retained, and transmitted. Inferences result in different privacy concerns that call for different governance mechanisms: misrepresentative inferences need correction or retraction, accurate but exposing inferences need restriction, and secondary uses need per-recipient and per-purpose approval. Users need ways to inspect what evidence produced an inference, decide whether it should persist, and control whether it travels beyond the context in which it was generated. Inference governance tools should distinguish between generation, retention, and transmission, and should let users act on the specific inferences a system has drawn about them throughout this lifecycle.


\bibliographystyle{ACM-Reference-Format}
\bibliography{references}

@article{barbosa2021design,
  title={Who am I? A design probe exploring real-time transparency about online and offline user profiling underlying targeted ads},
  author={Barbosa, Nat{\~a} M and Wang, Gang and Ur, Blase and Wang, Yang},
  journal={Proceedings of the ACM on Interactive, Mobile, Wearable and Ubiquitous Technologies},
  volume={5},
  number={3},
  pages={1--32},
  year={2021},
  publisher={ACM New York, NY, USA}
}

@article{mireshghallah2025cimemories,
  title={Cimemories: A compositional benchmark for contextual integrity of persistent memory in llms},
  author={Mireshghallah, Niloofar and Mangaokar, Neal and Kokhlikyan, Narine and Zharmagambetov, Arman and Zaheer, Manzil and Mahloujifar, Saeed and Chaudhuri, Kamalika},
  journal={arXiv preprint arXiv:2511.14937},
  year={2025}
}

@inproceedings{englehardt2016online,
  title={Online tracking: A 1-million-site measurement and analysis},
  author={Englehardt, Steven and Narayanan, Arvind},
  booktitle={Proceedings of the 2016 ACM SIGSAC conference on computer and communications security},
  pages={1388--1401},
  year={2016}
}

@inproceedings{gomez2023sensitive,
  title={What is sensitive about (sensitive) data? Characterizing sensitivity and intimacy with Google assistant users},
  author={G{\'o}mez Ortega, Alejandra and Bourgeois, Jacky and Kortuem, Gerd},
  booktitle={Proceedings of the 2023 CHI Conference on Human Factors in Computing Systems},
  pages={1--16},
  year={2023}
}

@inproceedings{leon2013matters,
  title={What matters to users? Factors that affect users' willingness to share information with online advertisers},
  author={Leon, Pedro Giovanni and Ur, Blase and Wang, Yang and Sleeper, Manya and Balebako, Rebecca and Shay, Richard and Bauer, Lujo and Christodorescu, Mihai and Cranor, Lorrie Faith},
  booktitle={Proceedings of the ninth symposium on usable privacy and security},
  pages={1--12},
  year={2013}
}

@inproceedings{wang2011regretted,
  title={``I regretted the minute I pressed share'': A qualitative study of regrets on Facebook},
  author={Wang, Yang and Norcie, Gregory and Komanduri, Saranga and Acquisti, Alessandro and Leon, Pedro Giovanni and Cranor, Lorrie Faith},
  booktitle={Proceedings of the seventh symposium on usable privacy and security},
  pages={1--16},
  year={2011}
}

@article{mireshghallah2025position,
  title={Position: Privacy Is Not Just Memorization!},
  author={Mireshghallah, Niloofar and Li, Tianshi},
  journal={arXiv preprint arXiv:2510.01645},
  year={2025}
}

@article{mireshghallah2024trust,
  title={Trust no bot: Discovering personal disclosures in human-llm conversations in the wild},
  author={Mireshghallah, Niloofar and Antoniak, Maria and More, Yash and Choi, Yejin and Farnadi, Golnoosh},
  journal={arXiv preprint arXiv:2407.11438},
  year={2024}
}

@article{mireshghallah2023can,
  title={Can llms keep a secret? testing privacy implications of language models via contextual integrity theory},
  author={Mireshghallah, Niloofar and Kim, Hyunwoo and Zhou, Xuhui and Tsvetkov, Yulia and Sap, Maarten and Shokri, Reza and Choi, Yejin},
  journal={arXiv preprint arXiv:2310.17884},
  year={2023}
}

@article{nissenbaum2004privacy,
  title={Privacy as contextual integrity},
  author={Nissenbaum, Helen},
  journal={Wash. L. Rev.},
  volume={79},
  pages={119},
  year={2004},
  publisher={HeinOnline}
}

@article{tomekcce2024private,
  title={Private attribute inference from images with vision-language models},
  author={T{\"o}mek{\c{c}}e, Batuhan and Vero, Mark and Staab, Robin and Vechev, Martin},
  journal={Advances in Neural Information Processing Systems},
  volume={37},
  pages={103619--103651},
  year={2024}
}

@inproceedings{schaub2016watching,
  title={Watching them watching me: Browser extensions’ impact on user privacy awareness and concern},
  author={Schaub, Florian and Marella, Aditya and Kalvani, Pranshu and Ur, Blase and Pan, Chao and Forney, Emily and Cranor, Lorrie Faith},
  booktitle={NDSS workshop on usable security},
  volume={10},
  year={2016}
}

@article{staab2023beyond,
  title={Beyond memorization: Violating privacy via inference with large language models},
  author={Staab, Robin and Vero, Mark and Balunovi{\'c}, Mislav and Vechev, Martin},
  journal={arXiv preprint arXiv:2310.07298},
  year={2023}
}

@inproceedings{tran2025understanding,
  title={Understanding Privacy Norms Around LLM-Based Chatbots: A Contextual Integrity Perspective},
  author={Tran, Sarah and Lu, Hongfan and Slaughter, Isaac and Herman, Bernease and Dangol, Aayushi and Fu, Yue and Chen, Lufei and Gebreyohannes, Biniyam and Howe, Bill and Hiniker, Alexis and others},
  booktitle={Proceedings of the AAAI/ACM Conference on AI, Ethics, and Society},
  volume={8},
  number={3},
  pages={2522--2534},
  year={2025}
}

@inproceedings{weinshel2019oh,
  title={Oh, the places you've been! User reactions to longitudinal transparency about third-party web tracking and inferencing},
  author={Weinshel, Ben and Wei, Miranda and Mondal, Mainack and Choi, Euirim and Shan, Shawn and Dolin, Claire and Mazurek, Michelle L and Ur, Blase},
  booktitle={Proceedings of the 2019 ACM SIGSAC Conference on Computer and Communications Security},
  pages={149--166},
  year={2019}
}

@inproceedings{zhang2024s,
  title={“It's a Fair Game”, or Is It? Examining How Users Navigate Disclosure Risks and Benefits When Using LLM-Based Conversational Agents},
  author={Zhang, Zhiping and Jia, Michelle and Lee, Hao-Ping and Yao, Bingsheng and Das, Sauvik and Lerner, Ada and Wang, Dakuo and Li, Tianshi},
  booktitle={Proceedings of the 2024 CHI Conference on Human Factors in Computing Systems},
  pages={1--26},
  year={2024}
}

@inproceedings{kacsmar2023comprehension,
  title={Comprehension from chaos: Towards informed consent for private computation},
  author={Kacsmar, Bailey and Duddu, Vasisht and Tilbury, Kyle and Ur, Blase and Kerschbaum, Florian},
  booktitle={Proceedings of the 2023 ACM SIGSAC Conference on Computer and Communications Security},
  pages={210--224},
  year={2023}
}

@inproceedings{lee2024deepfakes,
  title={Deepfakes, phrenology, surveillance, and more! a taxonomy of ai privacy risks},
  author={Lee, Hao-Ping and Yang, Yu-Ju and Von Davier, Thomas Serban and Forlizzi, Jodi and Das, Sauvik},
  booktitle={Proceedings of the 2024 CHI Conference on Human Factors in Computing Systems},
  pages={1--19},
  year={2024}
}

@inproceedings{shaikh2025creating,
  title={Creating general user models from computer use},
  author={Shaikh, Omar and Sapkota, Shardul and Rizvi, Shan and Horvitz, Eric and Park, Joon Sung and Yang, Diyi and Bernstein, Michael S},
  booktitle={Proceedings of the 38th Annual ACM Symposium on User Interface Software and Technology},
  pages={1--23},
  year={2025}
}

@article{lan2025contextual,
  title={Contextual integrity in LLMs via reasoning and reinforcement learning},
  author={Lan, Guangchen and Inan, Huseyin A and Abdelnabi, Sahar and Kulkarni, Janardhan and Wutschitz, Lukas and Shokri, Reza and Brinton, Christopher G and Sim, Robert},
  journal={arXiv preprint arXiv:2506.04245},
  year={2025}
}

@incollection{brusilovsky2007user,
  title={User models for adaptive hypermedia and adaptive educational systems},
  author={Brusilovsky, Peter and Mill{\'a}n, Eva},
  booktitle={The adaptive web: methods and strategies of web personalization},
  pages={3--53},
  year={2007},
  publisher={Springer}
}

@article{kobsa2007generic,
  title={Generic user modeling systems},
  author={Kobsa, Alfred},
  journal={The adaptive web: Methods and strategies of web personalization},
  pages={136--154},
  year={2007},
  publisher={Springer}
}

@inproceedings{ur2012smart,
  title={Smart, useful, scary, creepy: perceptions of online behavioral advertising},
  author={Ur, Blase and Leon, Pedro Giovanni and Cranor, Lorrie Faith and Shay, Richard and Wang, Yang},
  booktitle={proceedings of the eighth symposium on usable privacy and security},
  pages={1--15},
  year={2012}
}

@article{tajfel2001integrative,
  title={An integrative theory of intergroup conflict},
  author={Tajfel, Henri and Turner, John and Austin, William G and Worchel, Stephen and others},
  journal={Intergroup relations: Essential readings},
  pages={94--109},
  year={2001},
  publisher={Psychology Press}
}

@book{costa2000neo,
  title={Neo Personality Inventory.},
  author={Costa Jr, Paul T and McCrae, Robert R},
  year={2000},
  publisher={American Psychological Association}
}

@article{wood2007new,
  title={A new look at habits and the habit-goal interface.},
  author={Wood, Wendy and Neal, David T},
  journal={Psychological review},
  volume={114},
  number={4},
  pages={843},
  year={2007},
  publisher={American Psychological Association}
}

@article{tan2026memsifter,
  title={MemSifter: Offloading LLM Memory Retrieval via Outcome-Driven Proxy Reasoning},
  author={Tan, Jiejun and Dou, Zhicheng and Zhang, Liancheng and Hu, Yuyang and Cheng, Yiruo and Wen, Ji-Rong},
  journal={arXiv preprint arXiv:2603.03379},
  year={2026}
}

@article{thompson2007development,
  title={Development and validation of an internationally reliable short-form of the positive and negative affect schedule (PANAS)},
  author={Thompson, Edmund R},
  journal={Journal of cross-cultural psychology},
  volume={38},
  number={2},
  pages={227--242},
  year={2007},
  publisher={Sage Publications Sage CA: Thousand Oaks, CA}
}

@article{malhotra2004internet,
  title={Internet users' information privacy concerns (IUIPC): The construct, the scale, and a causal model},
  author={Malhotra, Naresh K and Kim, Sung S and Agarwal, James},
  journal={Information systems research},
  volume={15},
  number={4},
  pages={336--355},
  year={2004},
  publisher={Informs}
}

@article{gross2021validity,
  title={Validity and reliability of the scale internet users’ information privacy concerns (iuipc)},
  author={Gro{\ss}, Thomas},
  journal={Proceedings on Privacy Enhancing Technologies},
  year={2021}
}

@article{braun2006using,
  title={Using thematic analysis in psychology},
  author={Braun, Virginia and Clarke, Victoria},
  journal={Qualitative research in psychology},
  volume={3},
  number={2},
  pages={77--101},
  year={2006},
  publisher={Taylor \& Francis}
}

@inproceedings{caine2016local,
  title={Local standards for sample size at CHI},
  author={Caine, Kelly},
  booktitle={Proceedings of the 2016 CHI conference on human factors in computing systems},
  pages={981--992},
  year={2016}
}

@article{guest2006many,
  title={How many interviews are enough? An experiment with data saturation and variability},
  author={Guest, Greg and Bunce, Arwen and Johnson, Laura},
  journal={Field methods},
  volume={18},
  number={1},
  pages={59--82},
  year={2006},
  publisher={Sage Publications Sage CA: Thousand Oaks, CA}
}

@article{hennink2017code,
  title={Code saturation versus meaning saturation: how many interviews are enough?},
  author={Hennink, Monique M and Kaiser, Bonnie N and Marconi, Vincent C},
  journal={Qualitative health research},
  volume={27},
  number={4},
  pages={591--608},
  year={2017},
  publisher={Sage Publications Sage CA: Los Angeles, CA}
}

@inproceedings{zhang2025understanding,
  title={Understanding Users' Privacy Perceptions Towards LLM's RAG-based Memory},
  author={Zhang, Shuning and Ma, Rongjun and Ma, Ying and Li, Shixuan and Xu, Yiqun and Yi, Xin and Li, Hewu},
  booktitle={Proceedings of the 2025 Workshop on Human-Centered AI Privacy and Security},
  pages={10--19},
  year={2025}
}

@inproceedings{chen2026relational,
  title={Relational Gains, Privacy Strains: Exploring Users' Perceptions and Experiences with ChatGPT's Memory Feature},
  author={Chen, Cheng and Molina, Maria D and Liao, Mengqi and Cho Snyder, Eugene},
  booktitle={Proceedings of the 2026 CHI Conference on Human Factors in Computing Systems},
  pages={1--17},
  year={2026}
}

@inproceedings{mekioussa2026towards,
  title={Towards Usable, Privacy Respecting Long-Term Memory for LLM-based Conversational Agents},
  author={Mekioussa Malki, Lisa},
  booktitle={Proceedings of the Extended Abstracts of the 2026 CHI Conference on Human Factors in Computing Systems},
  pages={1--5},
  year={2026}
}

@article{wang2025beyond,
  title={Beyond PII: How users attempt to estimate and mitigate implicit LLM inference},
  author={Wang, Synthia and Peddinti, Sai Teja and Taft, Nina and Feamster, Nick},
  journal={arXiv preprint arXiv:2509.12152},
  year={2025}
}

@article{zhang2024ghost,
  title={"Ghost of the past": identifying and resolving privacy leakage from LLM's memory through proactive user interaction},
  author={Zhang, Shuning and Ye, Lyumanshan and Yi, Xin and Tang, Jingyu and Shui, Bo and Xing, Haobin and Liu, Pengfei and Li, Hewu},
  journal={arXiv preprint arXiv:2410.14931},
  year={2024}
}

@inproceedings{lam2026just,
  title={Just-In-Time Objectives: A General Approach for Specialized AI Interactions},
  author={Lam, Michelle S and Shaikh, Omar and Xu, Hallie and Guo, Alice and Yang, Diyi and Heer, Jeffrey and Landay, James A and Bernstein, Michael S},
  booktitle={Proceedings of the 2026 CHI Conference on Human Factors in Computing Systems},
  pages={1--26},
  year={2026}
}

@article{shaikh2026learning,
  title={Learning Next Action Predictors from Human-Computer Interaction},
  author={Shaikh, Omar and Teutschbein, Valentin and Gandhi, Kanishk and Chi, Yikun and Haber, Nick and Robinson, Thomas and Ram, Nilam and Reeves, Byron and Yang, Sherry and Bernstein, Michael S and others},
  journal={arXiv preprint arXiv:2603.05923},
  year={2026}
}

@misc{openai_chatgpt_ads,
  title={Our approach to advertising and expanding access to {ChatGPT}},
  author={{OpenAI}},
  year={2026},
  month={January},
  howpublished={\url{https://openai.com/index/our-approach-to-advertising-and-expanding-access/}}
}

@misc{anthropic_mcp,
  title={Introducing the {Model Context Protocol}},
  author={{Anthropic}},
  year={2024},
  month={November},
  howpublished={\url{https://www.anthropic.com/news/model-context-protocol}}
}

@misc{openai_mcp,
  title={New tools and features in the {Responses API}},
  author={{OpenAI}},
  year={2025},
  howpublished={\url{https://openai.com/index/new-tools-and-features-in-the-responses-api/}}
}

@misc{openclaw,
  title={{OpenClaw}: Open-source autonomous {AI} agent framework},
  author={{OpenClaw}},
  year={2026},
  howpublished={\url{https://docs.openclaw.ai/}}
}

@inproceedings{maes1993learning,
  title={Learning interface agents},
  author={Maes, Pattie and Kozierok, Robyn},
  booktitle={AAAI},
  volume={93},
  pages={459--465},
  year={1993}
}

@article{shneiderman1997direct,
  title={Direct manipulation vs. interface agents},
  author={Shneiderman, Ben and Maes, Pattie},
  journal={interactions},
  volume={4},
  number={6},
  pages={42--61},
  year={1997},
  publisher={ACM New York, NY, USA}
}

@article{lieberman1995letizia,
  title={Letizia: An agent that assists web browsing},
  author={Lieberman, Henry and others},
  journal={IJCAI (1)},
  volume={1995},
  pages={924--929},
  year={1995}
}

@inproceedings{farke2021privacy,
  title={Are privacy dashboards good for end users? Evaluating user perceptions and reactions to Google's My Activity},
  author={Farke, Florian M and Balash, David G and Golla, Maximilian and D{\"u}rmuth, Markus and Aviv, Adam J},
  booktitle={30th USENIX Security Symposium (USENIX Security 21)},
  pages={483--500},
  year={2021}
}

@inproceedings{zhou2025rescriber,
  title={Rescriber: Smaller-LLM-powered user-led data minimization for LLM-based chatbots},
  author={Zhou, Jijie and Xu, Eryue and Wu, Yaoyao and Li, Tianshi},
  booktitle={Proceedings of the 2025 CHI Conference on Human Factors in Computing Systems},
  pages={1--28},
  year={2025}
}

@article{zhang2024adanonymizer,
  title={Adanonymizer: interactively navigating and balancing the duality of privacy and output performance in human-LLM interaction},
  author={Zhang, Shuning and Yi, Xin and Xing, Haobin and Ye, Lyumanshan and Hu, Yongquan and Li, Hewu},
  journal={arXiv preprint arXiv:2410.15044},
  year={2024}
}

@inproceedings{asthana2024know,
  title={" I know even if you don't tell me": Understanding Users' Privacy Preferences Regarding AI-based Inferences of Sensitive Information for Personalization},
  author={Asthana, Sumit and Im, Jane and Chen, Zhe and Banovic, Nikola},
  booktitle={Proceedings of the 2024 CHI Conference on Human Factors in Computing Systems},
  pages={1--21},
  year={2024}
}

@article{reitinger2024does,
  title={What does it mean to be creepy? Responses to visualizations of personal browsing activity, online tracking, and targeted ads},
  author={Reitinger, Nathan and Wen, Bruce and Mazurek, Michelle L and Ur, Blase},
  journal={Proceedings on Privacy Enhancing Technologies},
  volume={2024},
  number={3},
  year={2024},
  publisher={Proceedings on Privacy Enhancing Technologies}
}

@article{farke2023does,
  title={How does connecting online activities to advertising inferences impact privacy perceptions?},
  author={Farke, Florian M and Balash, David G and Golla, Maximilian and Aviv, Adam J},
  journal={arXiv preprint arXiv:2312.13813},
  year={2023}
}

@article{buchi2023making,
  title={Making sense of algorithmic profiling: user perceptions on Facebook},
  author={B{\"u}chi, Moritz and Fosch-Villaronga, Eduard and Lutz, Christoph and Tam{\`o}-Larrieux, Aurelia and Velidi, Shruthi},
  journal={Information, Communication \& Society},
  volume={26},
  number={4},
  pages={809--825},
  year={2023},
  publisher={Taylor \& Francis}
}

@inproceedings{hautea2020s,
  title={'That's Not Me': Surprising Algorithmic Inferences},
  author={Hautea, Samantha and Munasinghe, Anjali and Rader, Emilee},
  booktitle={Extended abstracts of the 2020 CHI conference on human factors in computing systems},
  pages={1--7},
  year={2020}
}

@inproceedings{kronhardt2025all,
  title={All of That in 15 Minutes? Exploring Privacy Perceptions Across Cognitive Abilities via Ad-hoc LLM-Generated Profiles Inferred from Social Media Use},
  author={Kronhardt, Kirill and Hoffmann, Sebastian and Adelt, Fabian and Pascher, Max and Gerken, Jens},
  booktitle={Proceedings of the 27th International Conference on Multimodal Interaction},
  pages={164--172},
  year={2025}
}

@inproceedings{yun2026ai,
  title={AI and My Values: User Perceptions of LLMs' Ability to Extract, Embody, and Explain Human Values from Casual Conversations},
  author={Yun, Bhada and Su, Renn and Yi Wang, April},
  booktitle={Proceedings of the 2026 CHI Conference on Human Factors in Computing Systems},
  pages={1--38},
  year={2026}
}

@article{staufer2026human,
  title={Human-Centred LLM Privacy Audits: Findings and Frictions},
  author={Staufer, Dimitri and Morehouse, Kirsten and Hartmann, David and Berendt, Bettina},
  journal={arXiv preprint arXiv:2603.12094},
  year={2026}
}

@article{xu2026toward,
  title={Toward Personalized LLM-Powered Agents: Foundations, Evaluation, and Future Directions},
  author={Xu, Yue and Chen, Qi'an and Ma, Zizhan and Liu, Dongrui and Wang, Wenxuan and Wang, Xiting and Xiong, Li and Wang, Wenjie},
  journal={ACM Computing Surveys},
  year={2026}
}

@article{wu2024understanding,
  title={Understanding the Role of User Profile in the Personalization of Large Language Models},
  author={Wu, Bin and Shi, Zhengyan and Rahmani, Hossein A and Ramineni, Varsha and Yilmaz, Emine},
  journal={arXiv preprint arXiv:2406.17803},
  year={2024}
}

@article{peters2024large,
  title={Large language models can infer personality from free-form user interactions},
  author={Peters, Heinrich and Cerf, Moran and Matz, Sandra C},
  journal={arXiv preprint arXiv:2405.13052},
  year={2024}
}

@misc{chatgpt_memory,
  title={{Memory FAQ}},
  author={{OpenAI}},
  year={2026},
  month={April},
  howpublished={\url{https://help.openai.com/en/articles/8590148-memory-faq}},
  note={Accessed: April 29, 2026}
}

@misc{claude_memory,
  title={Bringing memory to {Claude}},
  author={{Anthropic}},
  year={2025},
  month={September},
  howpublished={\url{https://www.anthropic.com/news/memory}},
  note={Accessed: April 29, 2026}
}

@misc{gemini_memory,
  title={Save info and reference past chats in {Gemini Apps}},
  author={{Google}},
  year={2026},
  howpublished={\url{https://support.google.com/gemini/answer/15637730}},
  note={Accessed: April 29, 2026}
}

@article{fisher2026response,
  title={Response-Aware User Memory Selection for LLM Personalization},
  author={Fisher, Jillian and Neville, Jennifer and Park, Chan Young},
  journal={arXiv preprint arXiv:2604.14473},
  year={2026}
}

@inproceedings{yao2017folk,
  title={Folk models of online behavioral advertising},
  author={Yao, Yaxing and Lo Re, Davide and Wang, Yang},
  booktitle={Proceedings of the 2017 ACM Conference on Computer Supported Cooperative Work and Social Computing},
  pages={1957--1969},
  year={2017}
}

@inproceedings{kang2015my,
  title={$\{$“My$\}$ data just goes $\{$Everywhere:”$\}$ user mental models of the internet and implications for privacy and security},
  author={Kang, Ruogu and Dabbish, Laura and Fruchter, Nathaniel and Kiesler, Sara},
  booktitle={Eleventh symposium on usable privacy and security (SOUPS 2015)},
  pages={39--52},
  year={2015}
}

@inproceedings{lee2023and,
  title={When and why do people want ad targeting explanations? Evidence from a four-week, mixed-methods field study},
  author={Lee, Hao-Ping Hank and Logas, Jacob and Yang, Stephanie and Li, Zhouyu and Barbosa, Nata and Wang, Yang and Das, Sauvik},
  booktitle={2023 IEEE Symposium on Security and Privacy (SP)},
  pages={2903--2920},
  year={2023},
  organization={IEEE}
}

@inproceedings{he2018rethinking,
  title={Rethinking Access Control and Authentication for the Home Internet of Things ($\{$$\{$$\{$$\{$$\{$IoT$\}$$\}$$\}$$\}$$\}$)},
  author={He, Weijia and Golla, Maximilian and Padhi, Roshni and Ofek, Jordan and D{\"u}rmuth, Markus and Fernandes, Earlence and Ur, Blase},
  booktitle={27th USENIX Security Symposium (USENIX Security 18)},
  pages={255--272},
  year={2018}
}

@inproceedings{naeini2017privacy,
  title={Privacy expectations and preferences in an $\{$IoT$\}$ world},
  author={Naeini, Pardis Emami and Bhagavatula, Sruti and Habib, Hana and Degeling, Martin and Bauer, Lujo and Cranor, Lorrie Faith and Sadeh, Norman},
  booktitle={Thirteenth symposium on usable privacy and security (SOUPS 2017)},
  pages={399--412},
  year={2017}
}

@article{wu2025modeling,
  title={Modeling End-User Affective Discomfort With Mobile App Permissions Across Physical Contexts},
  author={Wu, Yuxi and Logas, Jacob and Ponda, Devansh Jatin and Haines, Julia and Li, Jiaming and Nichols, Jeffrey and Edwards, W Keith and Das, Sauvik},
  year={2025},
  publisher={Internet Society}
}

@inproceedings{herley2009so,
  title={So long, and no thanks for the externalities: the rational rejection of security advice by users},
  author={Herley, Cormac},
  booktitle={Proceedings of the 2009 workshop on New security paradigms workshop},
  pages={133--144},
  year={2009}
}

@inproceedings{reeder2007usability,
  title={Usability challenges in security and privacy policy-authoring interfaces},
  author={Reeder, Robert W and Karat, Clare-Marie and Karat, John and Brodie, Carolyn},
  booktitle={IFIP Conference on Human-Computer Interaction},
  pages={141--155},
  year={2007},
  organization={Springer}
}

@inproceedings{im2023less,
  title={Less is not more: Improving findability and actionability of privacy controls for online behavioral advertising},
  author={Im, Jane and Wang, Ruiyi and Lyu, Weikun and Cook, Nick and Habib, Hana and Cranor, Lorrie Faith and Banovic, Nikola and Schaub, Florian},
  booktitle={Proceedings of the 2023 CHI Conference on Human Factors in Computing Systems},
  pages={1--33},
  year={2023}
}

\appendix 

\section*{Open Science}
\label{app:open_science}
We provide an anonymous repository for double-blind review containing the artifacts needed to evaluate the Reflective Layer system and analysis workflow.\footnote{\url{https://anonymous.4open.science/r/reflective-layer-ccs/README.md}} The primary artifact is the Reflective Layer web application, which includes the participant-facing interface, provenance views, inference-generation pipeline, prompts/templates, model/provider/version and configuration details, dependency files, and environment/run instructions. We also provide separate synthetic ChatGPT-history exports, mock inferences, cached synthetic pipeline outputs, documentation, the qualitative codebook/coding schema, anonymized aggregate rating data, and analysis scripts or notebooks used to generate reported tables and figures.

We do not release raw participant ChatGPT exports, interview transcripts, recordings, sketches, or other participant-level private data because these materials contain sensitive and potentially identifying personal information and were not consented for public release. Instead, reviewers receive the synthetic and aggregate artifacts above, and the paper appendix provides implementation details (Appendix~\ref{app:implementation}), study instruments and interview/sketching protocol (Appendix~\ref{app:study_protocol}), quantitative-analysis details and the pipeline funnel (Appendix~\ref{app:analysis}), and additional results tables and figures (Appendix~\ref{app:additional_results}).

\section*{Ethical Considerations}
\label{app:ethical_considerations}

This study involved human participants and potentially sensitive personal data, because the Reflective Layer surfaced inferences drawn from participants' own ChatGPT conversation histories. The study protocol was reviewed and approved by our institution's IRB, and all participants provided informed consent before participating. Participants were compensated for their time and were told that they could stop the session, skip questions, or decline to discuss any surfaced inference.

The benefit of the study is to understand how users perceive, evaluate, and want to govern personal information inferred by LLM systems. The main risks were emotional discomfort, surprise self-disclosure, incorrect or stigmatizing inferences, reprocessing of historical chats, inadvertent disclosure during interviews, and exposure to third-party API processing. We judged these risks acceptable only with safeguards that minimized disclosure to the research team and preserved participant control over what was shared.

The study recruited ChatGPT users, so the source material processed by the Reflective Layer consisted of conversations that participants had already chosen to enter into OpenAI's ChatGPT service before the study. After consent, the research interface reprocessed only the bounded participant-selected subset described in Appendix~\ref{app:implementation} through OpenAI's API to generate inferences for the participant to inspect; the complete export was not sent wholesale. These API requests were transmitted over HTTPS, and participants were told before consenting that selected chat text would be processed through OpenAI's API. The study used the API project settings available to the research team to prevent API inputs and outputs from being used for model training; any provider-side logging or retention for abuse monitoring was governed by the API provider's terms rather than by our research data store.

We designed the protocol to minimize what participants had to disclose to researchers. The full set of conversations and generated inferences remained within the study web application during the session. Participants were asked to share their screen only during the introduction to the study and system interface; once the introduction ended and before participants began exploring the generated inferences, screen sharing was not required and could be stopped at any time without consequence. Researchers did not collect participants' raw chat histories, did not collect any inference unless the participant explicitly shared it, and did not require participants to directly share or discuss surfaced inferences they were not comfortable sharing. During exploration, survey, and interview activities, participants could respond at the level of their reaction or reasoning without revealing the underlying chat content or the full text of an inference.

Inspecting personal inferences can surface uncomfortable or unexpected information, so the session script emphasized participant control, voluntary disclosure, and debriefing. Participants were reminded that there were no right or wrong answers, that they could skip sensitive items, stop at any time, or withdraw after data sharing to the extent described in the consent materials. Shared inferences, survey responses, recordings, transcripts, and sketches were retained only for research analysis, stored in access-controlled research storage, and de-identified or reported only in aggregate or redacted form. These safeguards were intended to support the empirical study of LLM-derived personal inferences while limiting unnecessary exposure of participants' private conversational content.

The Reflective Layer also has dual-use risks: similar techniques could be repurposed for surveillance, profiling, advertising, or other unwanted inference about users. For that reason, our released artifacts exclude participant data and participant-specific inferences, and provide only research code, prompts, documentation, aggregate outputs, and synthetic example data.


\section{Reflective Layer Implementation}
\label{app:implementation}

The Reflective Layer is implemented as a web application. The uploaded source conversations came from recruited ChatGPT users and were histories that participants had already provided to OpenAI through ordinary ChatGPT use. After participants provided informed consent, model calls ran through OpenAI's API to reprocess a bounded subset of those same participant-selected histories for the study interface. This appendix records implementation details that support the pipeline description in the main text.

\emph{Conversation subset.} Running the full pipeline over an entire ChatGPT history is infeasible within the time and cost budget of an in-session study: heavy users can accumulate thousands of conversations, and each additional conversation multiplies both API calls and participant wait time. We therefore process a bounded subset of each participant's most recent ChatGPT conversations (up to 100 conversations; up to four user messages per conversation, drawn from the beginning and end). This captures the setup and resolution of each exchange---where users typically state intent and outcomes---while keeping the input small enough to complete end-to-end within the study session. Even at this scale, the pipeline produced enough inferences per participant to support the evaluation and selection tasks in our protocol.

\emph{Cross-conversation retrieval.} For each newly extracted observation, the pipeline retrieves related observations from prior conversations. This retrieval step uses lexical similarity while giving more weight to recent conversations, so that inference generation can draw on cross-conversation patterns without treating old and recent evidence as equally salient. The retrieved set grounds inference generation in more than a single exchange, enabling the system to identify recurring themes across the participant's conversation history.

\emph{Inference categories.} The six inference categories map onto enduring constructs in social identity, personality, and behavioral psychology~\cite{tajfel2001integrative, costa2000neo, wood2007new}, and build on prior usable privacy work showing the value of surfacing inferred advertising profiles in structured, inspectable forms~\cite{ur2012smart,barbosa2021design}. The categories are: \emph{identity} (demographics, professional identity, health status), \emph{psychometric traits} (personality, values, cognitive styles), \emph{interests} (hobbies, topics, expertise), \emph{behavioral patterns} (routines, habits, lifestyle), \emph{relationships} (family, networks, dynamics), and \emph{goals} (intentions, aspirations, objectives). An observation such as ``Works as a software engineer at Google,'' for instance, may yield the \emph{identity}-category inference ``Likely earns \$180--250K based on FAANG compensation norms''---distinct from, and derived from rather than restating the underlying observation.

\emph{Two-tier model configuration.} The pipeline uses two different classes of models. Observation extraction runs on a smaller, lower-cost model (GPT-4.1 Mini; temperature 0) because the task is structured and largely factual: extract titled records with source-message links. Inference generation and CI-based sensitivity scoring run on a stronger reasoning model (GPT-5.1 Chat; medium reasoning effort) because these stages require multi-observation reasoning, calibrated confidence, and CI-informed judgment; revision uses the same reasoning model to merge, weaken, or strengthen propositions. Cross-conversation retrieval uses lexical relevance with recency weighting (BM25 with temporal decay). To stay within model context limits and the study-session time budget, messages are truncated and conversations are chunked before model calls (message text capped at 4{,}000 characters before analysis; large conversations chunked at approximately 100k input tokens).

\emph{Sensitivity scoring.} For each inference, we prompt the model to reason through four Contextual Integrity parameters adapted from Lan et al.: (i)~\emph{information type} (e.g., health, financial, identity, relationship, or mundane); (ii)~\emph{social context} (e.g., professional, medical, intimate, casual); (iii)~\emph{norm expectation} (whether the user would expect this to be inferred); and (iv)~\emph{potential for harm} (whether the inference could enable embarrassment, discrimination, or other harm if disclosed). The reasoning is emitted alongside a continuous sensitivity score, with higher scores indicating inferences judged more private, unexpected, or potentially harmful.

\emph{Revision and deduplication.} Newly generated inferences are compared against existing ones, duplicates are merged, contradictions are resolved, and confidence scores are updated in light of corroborating or conflicting evidence. This stage mirrors GUM's \emph{Revise} logic and produces the final set of inferences displayed to participants.

\section{Study Protocol}
\label{app:study_protocol}

This appendix provides the exact survey items and interview prompts used in the study protocol.

\subsection{Recruitment and Sample Size Rationale}

Our sample size is consistent with established benchmarks for qualitative interview studies: a survey of user-study sample sizes in HCI places the modal interview study at twelve participants~\cite{caine2016local}, and other highly cited studies on sample sizes for qualitative research found that saturation is commonly reached within 12--24 interviews~\cite{guest2006many,hennink2017code}.

\subsection{Session Flow}

Each session lasted approximately 90 minutes. The researcher first oriented participants to the study structure and research probe, reminded them that they could skip sensitive items or stop at any time, and introduced the system interface while participants shared their screen. Participants were then asked to upload their ChatGPT conversation export into the study web application. Once the introduction ended and before participants began exploring the generated inferences, screen sharing was not required, and participants could stop sharing their screen with the researcher at any time without consequence. Participants explored the generated inferences for approximately 15--20 minutes across the graph, dashboard, and sensitivity-ranked views. They then selected six inferences---one from each category: identity, psychometric traits, interests, behavioral patterns, relationships, and goals. The system assigned six additional inferences by selecting the highest-sensitivity inference in each category that the participant had not selected, yielding twelve inferences for structured evaluation where possible.

Participants then completed the structured per-inference survey while thinking aloud when comfortable. The survey focused on the inference itself, with source messages shown for context. After all inference ratings, participants completed affective and background measures. The session ended with a semi-structured interview and sketching activity probing participants' reasoning about inference appropriateness, access-control boundaries, desired actions over inferences, and conditions for sharing.

\subsection{Per-Inference Evaluation Items}

For each selected inference, participants rated the following six items on a 5-point agreement scale.

\begin{table}[h]
  \centering
  \caption{Per-inference constructs and item wording.}
  \label{tab:per_inference_constructs}
  \begin{tabularx}{\columnwidth}{@{}lX@{}}
    \toprule
    Construct & Item wording \\
    \midrule
    Accuracy & ``This inference is accurate about me/my life.'' \\
    Salience & ``This inference represents something meaningful about me and/or my life.'' \\
    Usefulness & ``It would be valuable for the system to know this about me.'' \\
    Intrusiveness & ``This inference feels like a violation of my personal boundaries.'' \\
    Comfort & ``I am comfortable with the system having made this inference about me.'' \\
    Surprise & ``I did not expect the system to make this inference about me.'' \\
    \bottomrule
  \end{tabularx}
\end{table}

\subsection{Access-Control Items}

For each selected inference, participants rated the following recipient-use scenarios on a 5-point agreement scale.

\begin{itemize}
  \item \textbf{Conversational improvement:} ``I am comfortable with the AI platform (e.g., OpenAI) using this inference to improve the relevance and quality of future conversations.'' Examples shown to participants included adjusting future conversation content based on the inference and understanding conversational tone preferences based on the inference.
  \item \textbf{Advertising:} ``I am comfortable with advertisers using this inference to show me relevant advertisements.'' Examples shown to participants included presenting ads for products based on the inference and personalizing promotional content based on the inference.
  \item \textbf{Action-taking:} ``I am comfortable with other third-party tools and services using this inference to personalize actions taken on my behalf.'' Examples shown to participants included considering the inference when drafting an email for the user and referring to the inference when scheduling an appointment on the user's behalf.
\end{itemize}

\subsection{Post-Exposure and Background Measures}

After participants completed the per-inference ratings, they completed the PANAS-10 short form to capture their affective response to seeing the surfaced inferences. The positive-affect items were interested, alert, inspired, enthusiastic, and attentive; the negative-affect items were distressed, upset, nervous, afraid, and irritable. Participants were then asked to explain the affective items they endorsed most and least strongly.

At the end of the session, participants reported demographics, LLM usage patterns, and privacy attitudes using an adapted eight-item IUIPC scale. The LLM usage questions covered frequency of use, typical purposes, and types of information typically shared. The IUIPC items covered control, awareness, and collection concerns, with ``online companies'' adapted to ``online services'' for the AI/LLM context.

\subsection{Interview and Sketching Prompts}

The semi-structured interview began with open-ended reflection prompts:

\begin{itemize}
  \item ``What stood out to you about the inferences you saw?''
  \item ``Were there any that felt especially meaningful, surprising, or uncomfortable? Why?''
  \item ``How did you decide whether an inference felt appropriate or intrusive?''
\end{itemize}

The sketching activity used the following core prompts:

\begin{itemize}
  \item ``Using this canvas, try to draw or organize how you'd want that kind of information handled. There's no right way to do this---use whatever structure makes sense to you.''
  \item ``Can you walk me through what you're drawing?''
  \item ``What's the difference between this area and that area?''
  \item ``Take each category and place it somewhere in what you've drawn. Tell me why each goes there.''
  \item ``Does it fit neatly, or is it hard to place? Why?''
\end{itemize}

Where time allowed, the researcher used follow-up probes:

\begin{itemize}
  \item ``If you could do anything about these inferences, what would you want to be able to do?''
  \item ``What's driving those preferences for you?''
  \item ``Beyond those, is there anything else that would affect whether this inference should be shared?''
\end{itemize}

\section{Analysis}
\label{app:analysis}
\subsection{Pipeline Funnel}

Table~\ref{tab:pipeline_funnel} summarizes how participants' uploaded histories were reduced to the inference-level ratings analyzed in the paper. Values are computed across the 18 non-pilot study participants unless otherwise noted.

\begin{table}[t]
  \centering
  \small
  \caption{Pipeline funnel from uploaded ChatGPT histories to final rated inferences ($N=18$ participants).}
  \label{tab:pipeline_funnel}
  \begin{tabularx}{\columnwidth}{@{}>{\raggedright\arraybackslash}Xrrll@{}}
    \toprule
    Stage & Mean & Median & Range & Total / notes \\
    \midrule
    User messages analyzed &
      208.1 & 214.5 & 52--322 & 3,746 total \\
    Observations generated &
      554.1 & 565 & 103--955 & 9,973 total \\
    Inferences generated &
      410.8 & 455.5 & 106--593 & 7,395 total \\
    Rated inferences &
      11.9 & 12 & 11--12 & 215 total \\
    \bottomrule
  \end{tabularx}
\end{table}

\subsection{Quantitative Analysis Details}

The quantitative analysis included two steps. First, descriptive statistics and inference-level correlations ($N=215$) surfaced direction and magnitude of each relationship, with pipeline metadata (category, confidence, Contextual Integrity-based sensitivity) retained for each rated inference. Second, we fit cumulative-link mixed models (CLMMs) to account for the ordinal five-point Likert outcomes and the nesting of inference-level observations within participants. Each CLMM used a logit link, adaptive Gaussian quadrature (nAGQ $=7$), and a random intercept for each participant to account for repeated measures. We fit both single-predictor and simultaneous-predictor specifications. The specific predictor--outcome pairings tested under CLMM were chosen to follow up on patterns surfaced by the qualitative analysis and the descriptive correlations. Coefficients in the CLMMs are reported on the log-odds scale: a positive $\beta$ means a unit increase in the predictor is associated with a shift toward higher Likert categories. CLMM significance for the primary construct-level contrasts supports inferential interpretation at this $N$; secondary findings that depend on smaller effects are reported descriptively as exploratory.

\begin{table}[!tbp]
  \centering
  \small
  \caption[Predictors of access-control comfort by recipient-purpose pairs]{Predictors of access-control comfort by recipient-purpose pairs. Coefficients are from simultaneous CLMMs for each recipient ($^{***}p<.001$, $^{**}p<.01$, $^{*}p<.05$).}
  \label{tab:per_recipient_predictors}
  \begin{tabularx}{\columnwidth}{@{}>{\raggedright\arraybackslash}Xccc@{}}
    \toprule
    Predictor & Platform & Advertisers & Action-taking \\
    \midrule
    Accuracy & +0.14 & +0.18 & +0.11 \\
    Usefulness & +0.89$^{***}$ & -0.01 & +0.41$^{*}$ \\
    Intrusiveness & -0.64$^{***}$ & -0.62$^{**}$ & -0.57$^{*}$ \\
    Surprise & -0.13 & -0.51$^{**}$ & -0.37$^{*}$ \\
    Salience & -0.13 & +0.29 & +0.03 \\
    \bottomrule
  \end{tabularx}
\end{table}

\section{Additional Results Figures and Tables}
\label{app:additional_results}

\begin{table*}[t]
  \centering
  \caption{Bridging how empirical patterns map to governance implications.}
  \label{tab:results_discussion_bridge}
  \begin{tabularx}{\textwidth}{@{}XXX@{}}
    \toprule
    Empirical pattern & Governance concern & Design implication \\
    \midrule
    Low accuracy + low comfort & Misrepresentation & Correct or retract the inference \\
    High accuracy + low usefulness & Exposure or overstep & Restrict use, or decay if the inference is context-bound \\
    Intrusive + surprising & Expectation violation & Show provenance and set expectations about inferential scope \\
    Platform comfort $>$ third-party comfort & Secondary use & Require per-recipient and per-purpose approval \\
    Staleness or calcification & Temporal or evidential mismatch & Add expiration rules and evidence thresholds \\
    \bottomrule
  \end{tabularx}
\end{table*}

\begin{figure}[p]
  \centering
  \includegraphics[width=\linewidth]{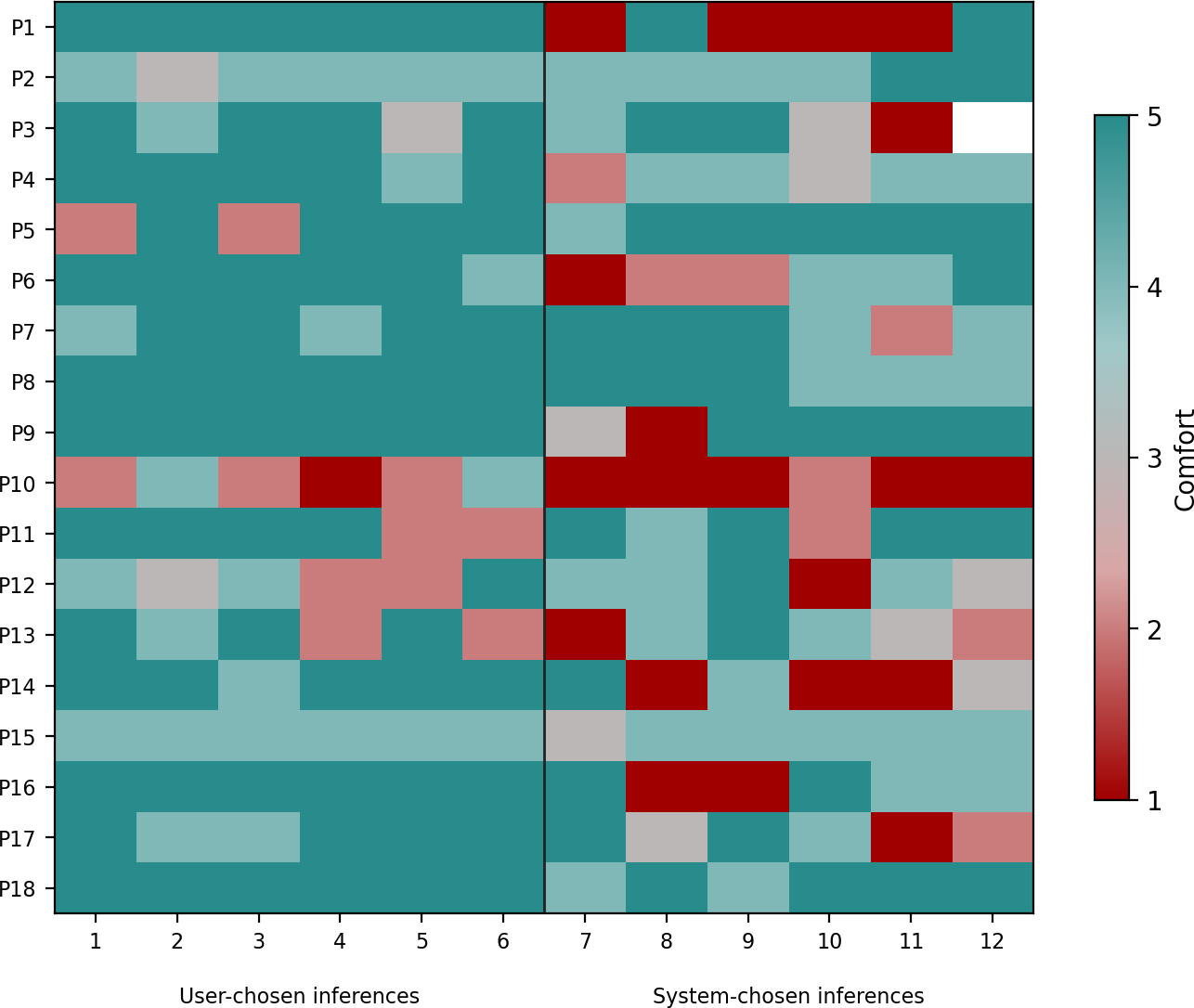}
  \caption{Comfort is not uniform within participants. Inferences are shown in the order participants rated them: the first six columns are participant-chosen inferences and the remaining columns are system-chosen inferences.}
  \label{fig:comfort-heatmap}
\end{figure}


Beyond differences in average comfort ratings across inference recipients, the factors that most strongly predicted comfort ratings also differed by recipient. When we modeled how participants' other rating dimensions correlated with access-control comfort, different dimensions carried the most weight for each recipient-purpose pair (Table~\ref{tab:per_recipient_predictors}).

\begin{figure}[p]
  \centering
  \includegraphics[width=\linewidth]{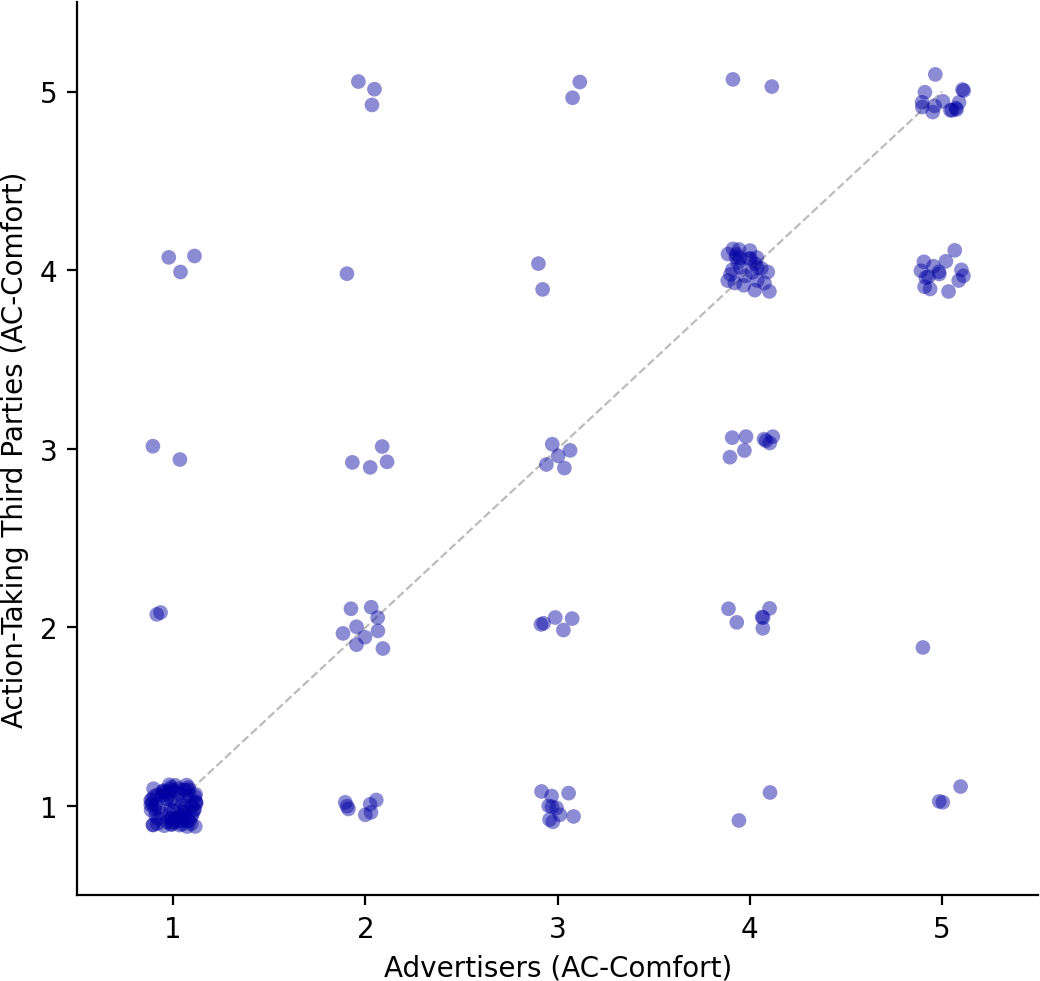}
  \caption{Advertiser and action-taking comfort track together. Each point is one inference's access-control comfort rating for advertiser use (showing relevant ads) and action-taking third-party use (personalizing actions on the user's behalf); the diagonal marks identical ratings. These two non-platform contexts were highly correlated (Spearman $\rho = 0.77$), while each correlated less strongly with platform use for improving future conversations ($\rho = 0.49$ for platform--advertisers; $\rho = 0.48$ for platform--action-taking).}
  \label{fig:advertiser-tools}
\end{figure}

\begin{figure}[p]
  \centering
  \includegraphics[width=\linewidth]{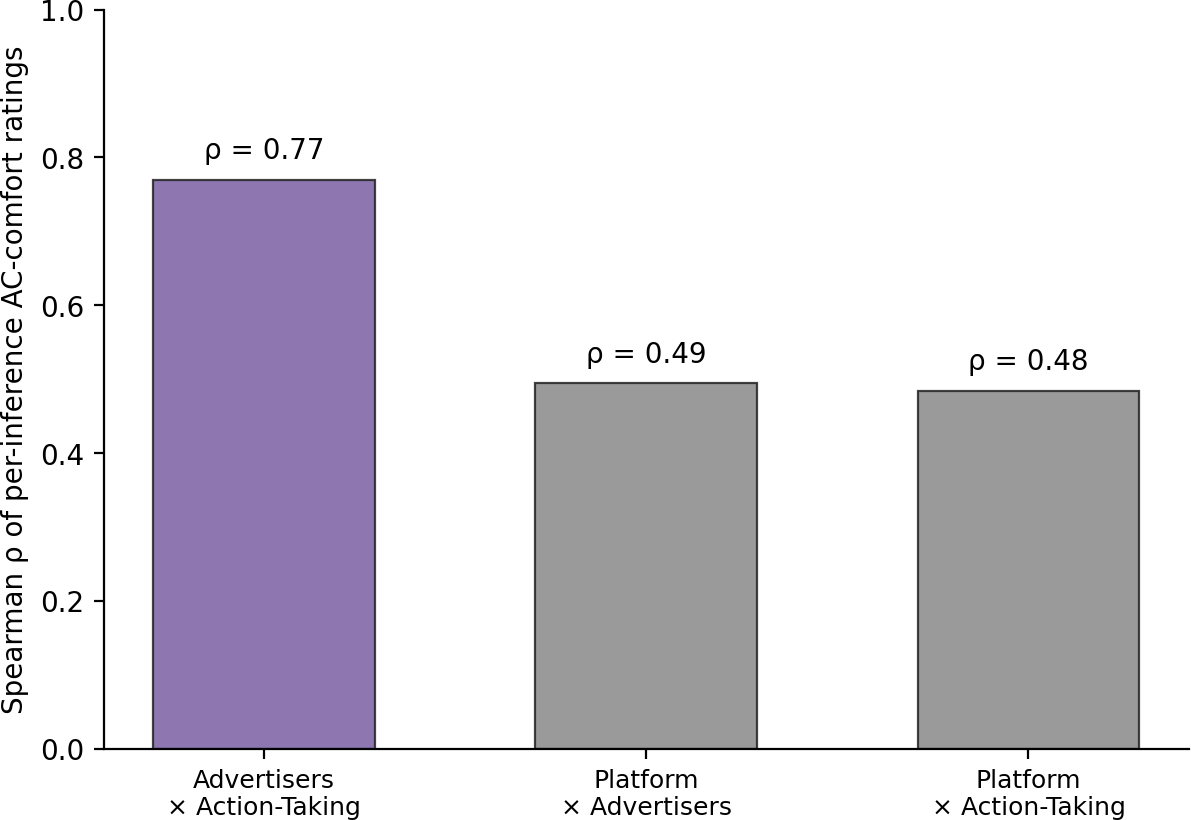}
  \caption{Advertiser--action-taking correlation compared with each context's correlation with platform access-control comfort. The platform scenario was using the inference to improve future conversations; the advertiser scenario was showing relevant ads; the action-taking scenario was personalizing actions on the user's behalf.}
  \label{fig:recipient-corr}
\end{figure}

\begin{figure}[p]
  \centering
  \includegraphics[width=\linewidth]{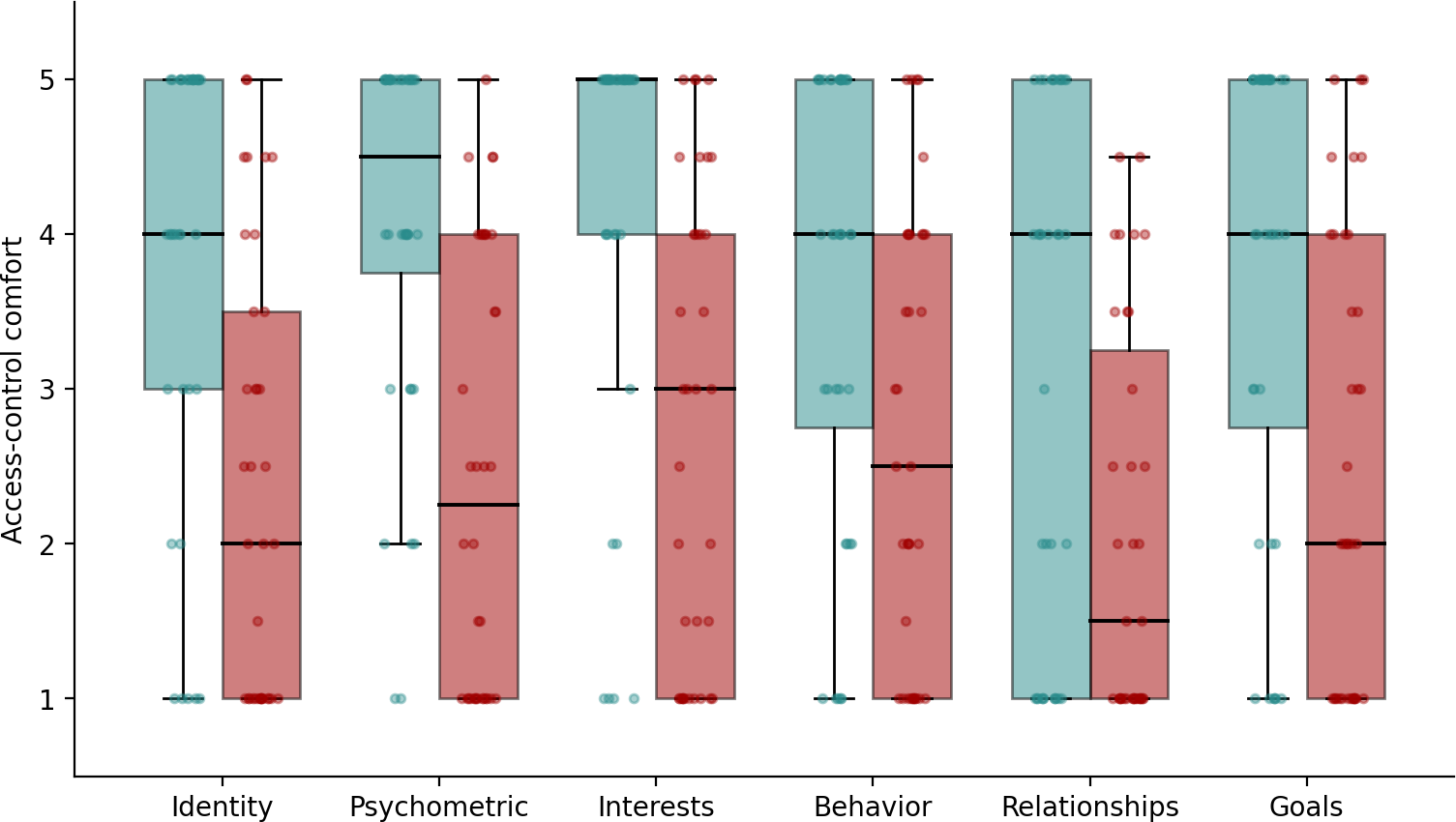}
  \caption{Within-category spread: category alone does not determine comfort. Green summarizes platform access-control comfort; red summarizes non-platform comfort, averaged across advertisers and action-taking third parties.}
  \label{fig:category-spread}
\end{figure}

\begin{figure}[p]
  \centering
  \includegraphics[width=\linewidth]{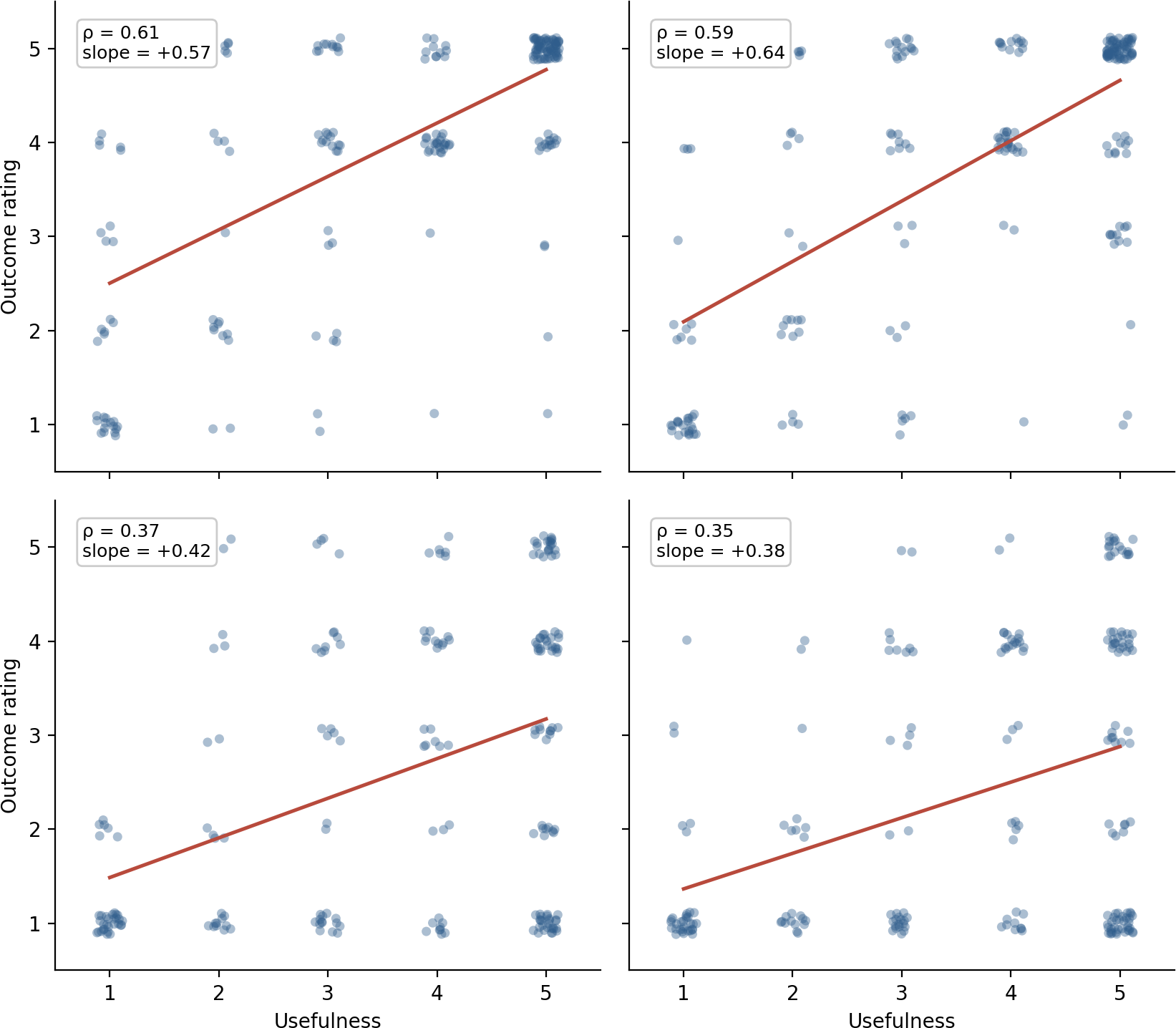}
  \caption{Usefulness carries less weight beyond the platform. Each panel shows participants' Usefulness rating ($x$) against one outcome rating ($y$), both on 5-point scales. ``First-order Comfort'' is comfort with the system having made the inference; the access-control panels ask about platform use to improve future conversations, advertiser use to show relevant ads, and third-party tool use to personalize actions on the user's behalf. Blue points = 215 per-inference ratings (jittered); red line = linear fit; $\rho$ = Spearman rank correlation.}
  \label{fig:usefulness-transfer}
\end{figure}

\begin{figure}[p]
  \centering
  \includegraphics[width=\linewidth]{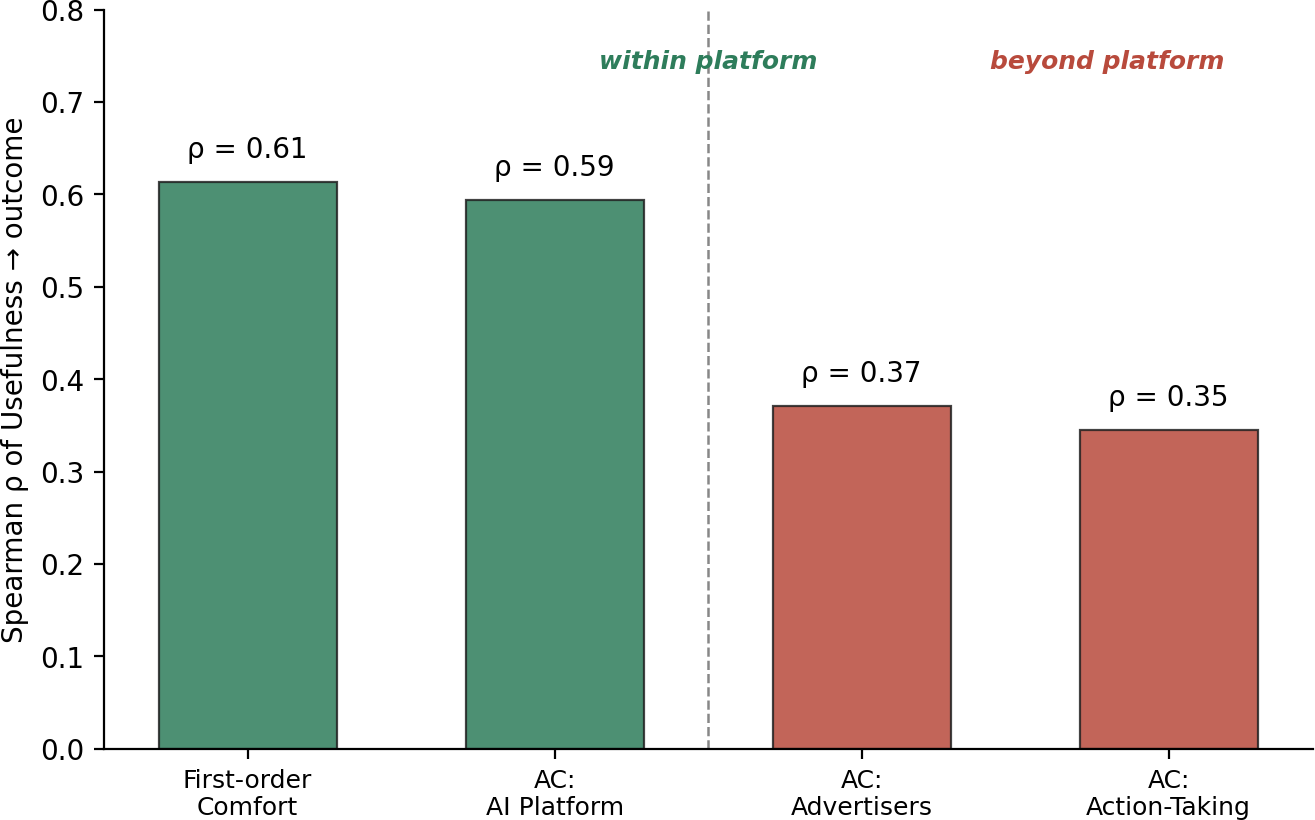}
  \caption{Attenuation in the association between perceived usefulness and comfort as inference use moves from the platform to non-platform recipients.}
  \label{fig:usefulness-attenuation}
\end{figure}

\end{document}